\documentclass[12pt]{article}
\usepackage{graphicx}
\usepackage{amsmath}
   \usepackage{ccaption}
    \usepackage[small,bf]{caption}
\begin{document}

\title{\textbf{Analysis of Coherence Properties of 3-rd Generation Synchrotron
Sources and Free-Electron Lasers}}
\author{I.A. Vartanyants\thanks{%
Corresponding author: Ivan.Vartaniants@desy.de} ~and A. Singer\\
 \small\textit{HASYLAB at DESY, Notkestr. 85, D-22607 Hamburg, Germany}}
\date{}
\maketitle


\begin{abstract}
A general theoretical approach based on the results of  statistical optics
is used for the analysis of the transverse coherence properties of 3-rd
generation synchrotron sources and x-ray free-electron lasers (XFEL).
Correlation properties of the wavefields are calculated at different
distances from an equivalent Gaussian Schell-model source. This model is
used to describe coherence properties of the five meter undulator source at
the synchrotron storage ring PETRA III. In the case of XFEL sources the
decomposition of the statistical fields into a sum of independently
propagating transverse modes is used for the analysis of the coherence
properties of these new sources. A detailed calculation is performed for
the parameters of the SASE1 undulator at the European XFEL. It is
demonstrated that only a few modes contribute significantly to the total
radiation field of that source.
\end{abstract}

\ 


\newpage
\section{ Introduction}

With the construction of 3-rd generation, hard x-ray synchrotron sources
(ESRF, APS and SPring8) with small source sizes and long distances from
source to sample (Fig. 1 (a)), new experiments that exploit the high
coherence properties of these x-ray beams have become feasible. New, high
brilliance, low emittance, hard x-ray synchrotron sources, such as PETRA
III, are under construction \cite{PETRA} and will provide an even higher
coherent photon flux for users. Experiments exploiting the coherence
properties of these x-ray beams become even more important with the
availability of 4-th generation x-ray sources - so-called x-ray
free-electron lasers (XFEL) (Fig. 1 (b)). These are presently in the
comissioning phase in the USA \cite{LCLS} and under construction in Japan
and Europe \cite{SCSS,XFEL}). Based on the self amplified spontaneous
emission (SASE) process \cite{SSY00}, they will provide ultrashort, coherent
x-ray pulses of unprecedentedly high brightness.

New areas of research have emerged that exploit the high coherence
properties of x-ray beams including x-ray photon correlation spectroscopy
(XPCS) (for a review see \cite{GZ04,L07}) and coherent x-ray diffractive
imaging (CXDI) \cite{MCK99,PWV06,CBM06,STB05,ANW08,TDM08}. In the former,
the dynamic properties of a system are studied and in the latter static
features are analyzed and a real space image of the sample can be obtained
by phase retrieval techniques \cite{F82,E03}. Similar ideas can be realized
through the use of newly emerging FEL sources. These experiments can be
further extended by the use of coherent femtosecond pulses of extremely high
intensity. The first demonstration experiments of the possibility of single
pulse \cite{CBB06,CHB07} and single pulse train \cite{MSR09} coherent
diffraction imaging were performed recently at the first FEL for extended
ultraviolet (XUV) wavelengths, the free-electron laser in Hamburg (FLASH) 
\cite{A07}. In the future, when x-ray FELs with a unique femtosecond time
structure will become available, this approach can even be used for
different applications in materials science \cite{VRM07}, the study of
dynamics \cite{GSG07} and biology including such exciting possibilities as
single molecule imaging \cite{NWS00}.

\begin{figure}[tbp]
\centering
	\includegraphics[width=0.9\textwidth]{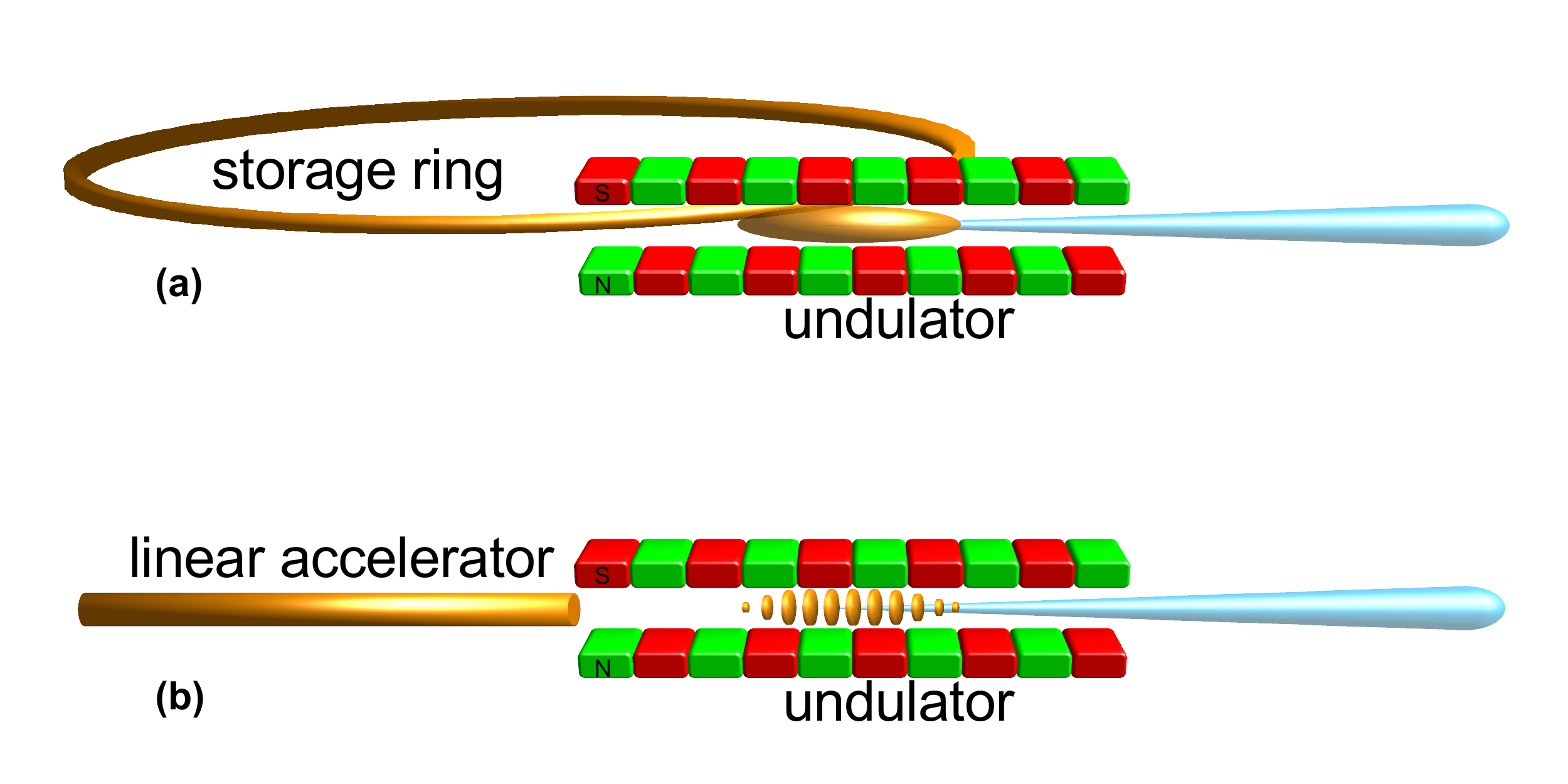} %
\caption{(a) Schematic view of the spontaneous radiation from the
undulator source, (b) Schematic view of the coherent radiation from the
free-electron laser based on the SASE principle.}
\end{figure}

From these perspectives it is clear that understanding the coherence
properties of beams emerging from 3-rd generation synchrotron sources and
new FELs is of vital importance for the scientific community. This includes
beamline scientists designing beamlines for coherent applications, or
experimentalists planning experiments with coherent beams. A beamline
scientist, for example, would like to estimate the performance of optics in
the partially coherent beam emerging from a synchrotron source. Nowadays,
several codes are used for simulation of x-ray propagation from a source
through the optical system of a beamline towards an experimental hutch. Most
of them are based either on the ray tracing approach (SHADOW \cite{SHADOW},
RAY \cite{RAY}, \textit{etc.}) which is, in fact, the geometrical optics
limit, or on the Fourier optics approach (PHASE \cite{PHASE}, \textit{etc.}%
). It will be shown in our paper that the first approach of ray tracing can
be safely used for synchrotron radiation in the horizontal direction, where
the radiation field is mostly incoherent, but can not be used in the
vertical direction, where the radiation from the undulator sources is highly
coherent. Unfortunately, the approach of Fourier optics in the description
of the scattering of synchrotron radiation in the vertical direction is also
limited when the radiation is not fully coherent but rather partially
coherent. Simulations performed for the XFEL sources at their saturation 
\cite{SSY08} suggest that these sources are not fully coherent, and as a
consequence, partial coherence effects must be carefully considered for them
as well. The contribution of partial coherence effects is also important in
phase retrieval as a reduced coherence of the incoming beam can produce
artifacts in reconstructed images \cite{VR01,VR03,WQP07}.

Different approaches may be used to analyze the coherence properties of the
synchrotron and XFEL sources. One is based on a detailed modeling of the
radiation process of ultrarelativistic particles in a storage ring or an
insertion device (see for e.g. \cite{THM98,GSS08}), or a detailed modeling
of the SASE process by performing calculations of nonlinear electromagnetic
equations at different conditions of operation (linear regime, saturation,
etc.) \cite{SSY08}. Another possible approach is based on the results of
statistical optics \cite{G85,MW95}, when the statistics of the radiation
wavefield is analyzed with the very general assumptions about the origin of
the radiators \cite{K86,HK94,C95}. An attractiveness of this approach is
based on the fact that with just a few parameters the coherence properties
of the beam can be described at different distances from the source. A
famous example of such an approach is the van Cittert-Zernike theorem \cite%
{G85,MW95} that predicts the coherence properties of an incoherent source at
any distance from that source. The most important assumption when deriving
this theorem is that the source is completely incoherent. This means that
there are no correlations between any two points of the source at any
separation between these two points. As such, it does not require a detailed
description of the physical origin of the radiation process of the source.
This theorem is widely applied for the estimate of the coherence properties
of 3-rd generation synchrotron sources. This is based on the assumption that
each electron in the electron bunch is radiating independently from another
electron, which means the radiation should be considered as incoherent.
However, this assumption, which is absolutely correct from the point of
radiation physics, brings us to a certain contradiction when a radiation
field at a large distance from a synchrotron source is analysed. The van
Cittert-Zernike theorem gives a realistic estimate of the coherence length
downstream from the source, but at the same time it predicts that such a
source should radiate in the solid angle of $4\pi $ as a completely
incoherent source. However, it is well known from electrodynamics (see for
e.g. \cite{LL}) that ultrarelativistic particles radiate mostly in the
forward direction in a narrow cone $\Delta \theta \sim \sqrt{1-v^{2}/c^{2}}$
around the direction of the velocity vector $\mathbf{v}$, where $v$ is the
velocity of the particle and $c$ is the speed of light. As a consequence,
typical x-ray beams produced by synchrotron radiation sources have a beam
character with a narrow cone and negligible radiation in all other
directions. The only way to solve this contradiction in the frame of
statistical optics is to assume a certain degree of coherence for a
synchrotron source. With this approach we will substitute a real synchrotron
source by an \textit{equivalent source} with a certain source size and, what
is especially important, a finite degree of coherence that will produce
wavefields with the same statistical properties as a radiation field from a
real source. With these general assumptions we will use the results of
statistical optics to calculate the statistical properties of beams emerging
from a new source, PETRA III, at different distances from the source.

We use a different approach for the description of the coherence properties
of the FELs. These highly coherent sources can be described with a finite
number of transverse modes. We will use a decomposition of the statistical
fields into a sum of independently propagating transverse modes and
calculate correlation fields at different distances from the source. The
source itself will be described by the same Gaussian functions, as in the
case of the synchrotron radiation, but clearly with different values of the
source parameters. Recently, we characterized the coherence properties of
FLASH using this method \cite{SVK08}. For the calculations in this paper we
will consider the concrete example of the European XFEL SASE1 undulator
source \cite{XFEL}. We will show that for this source only a few modes
contribute substantially to the total radiation field.

The paper is organized as follows. In the next section the basic equations
of statistical optics will be presented. In section three this general
approach will be applied for the characterization of the statistical
properties of the beams emerging from the 3-rd generation synchrotron
sources. In section four the coherent-mode representation of wavefields will
be used for the characterization of the coherence properties of the XFEL
beams. The paper ends with the conclusions and an outlook. In the Appendix we
compare results obtained in this paper with the theoretical approach
developed in Ref. \cite{GSS08} for calculation of coherence properties
of radiation emerging from the undulator sources.

\section{ Basic Equations}


\subsection{Correlation functions of wavefields}

The central concept in the theory of partial coherence is the so-called
mutual coherence function (MCF), $\Gamma (\mathbf{r}_{1},\mathbf{r}_{2};\tau
) $, that describes the correlations between two complex scalar\footnote{%
In the following, for simplicity, we consider only one polarization of the
x-ray field.} 
values of the electric field at different points $\mathbf{r}_{1}$ and $%
\mathbf{r}_{2}$ and at different times. It is defined as \cite{G85,MW95} 
\begin{equation}
\Gamma (\mathbf{r}_{1},\mathbf{r}_{2};\tau )=\left\langle E(\mathbf{r}%
_{1},t+\tau )E^{\ast }(\mathbf{r}_{2},t)\right\rangle _{T},  \label{2.1}
\end{equation}%
where $E(\mathbf{r}_{1},t+\tau )$ and $E(\mathbf{r}_{2},t)$ are the field
values at the points $\mathbf{r}_{1}$ and $\mathbf{r}_{2}$, $\tau $ is the
time delay, and brackets $\left\langle ...\right\rangle _{T}$ mean an
averaging over times $T$ much longer than the fluctuation time of the x-ray
field. It is also assumed that the radiation is ergodic and stationary. From
the definition of the MCF it follows that when two points coincide an
averaged intensity is given by 
\begin{equation}
<I(\mathbf{r})>=\Gamma (\mathbf{r},\mathbf{r};0)=\left\langle E(\mathbf{r}%
,t)E^{\ast }(\mathbf{r},t)\right\rangle _{T}.  \label{2.2}
\end{equation}%
It is usual to normalize the MCF as%
\begin{equation}
\gamma (\mathbf{r}_{1},\mathbf{r}_{2};\tau )=\frac{\Gamma (\mathbf{r}_{1},%
\mathbf{r}_{2};\tau )}{\sqrt{\left\langle I(\mathbf{r}_{1})\right\rangle }%
\sqrt{\left\langle I(\mathbf{r}_{2})\right\rangle }} , \label{2.3}
\end{equation}%
which is known as the complex degree of coherence. For all values of the
arguments $\mathbf{r}_{1},\mathbf{r}_{2}$ and $\tau $ the absolute value of
the complex degree of coherence $0\leq |\gamma (\mathbf{r}_{1},\mathbf{r}%
_{2};\tau )|\leq 1$.

Following Mandel \& Wolf \cite{MW95} the cross-spectral density (CSD)
function, $W(\mathbf{r}_{1},\mathbf{r}_{2},\omega )$, can be introduced.
According to its definition it forms a Fourier transform pair with the MCF
in the time-frequency domain%
\begin{equation}
W(\mathbf{r}_{1},\mathbf{r}_{2};\omega )=\int\limits_{-\infty }^{\infty
}\Gamma (\mathbf{r}_{1},\mathbf{r}_{2};\tau )e^{i\omega \tau }d\tau .
\label{2.4}
\end{equation}%
The cross-spectral density is a measure of correlation between the spectral
amplitudes of any particular frequency component $\omega $ of the light
vibrations at the points $\mathbf{r}_{1}$ and $\mathbf{r}_{2}$. When two
points $\mathbf{r}_{1}$ and $\mathbf{r}_{2}$ coincide, the CSD represents
the spectral density, $S(\mathbf{r},\omega )$, (or the power spectrum) of the
field%
\begin{equation}
S(\mathbf{r},\omega )=W(\mathbf{r},\mathbf{r};\omega ).  \label{2.6}
\end{equation}%
Introducing the normalized cross-spectral density function, $\mu (\mathbf{r}%
_{1},\mathbf{r}_{2};\omega )$, which is called the spectral degree of
coherence (SDC) at frequency $\omega $, we obtain 
\begin{equation}
\mu (\mathbf{r}_{1},\mathbf{r}_{2};\omega )=\frac{W(\mathbf{r}_{1},\mathbf{r}%
_{2};\omega )}{\sqrt{S(\mathbf{r}_{1},\omega )}\sqrt{S(\mathbf{r}_{2},\omega
)}},  \label{2.7}
\end{equation}%
where again, as for the case of the complex degree of coherence, the
following inequality is valid $0\leq |\mu (\mathbf{r}_{1},\mathbf{r}%
_{2};\omega )|\leq 1$.

In order to characterize the transverse coherence properties of the
wavefields by one number the degree of the transverse coherence can be
introduced as \cite{SSY08}%
\begin{equation}
\zeta (\omega )=\frac{\int |\mu (\mathbf{r}_{1},\mathbf{r}_{2};\omega
)|^{2}S(\mathbf{r}_{1};\omega )S(\mathbf{r}_{2};\omega )d\mathbf{r}_{1}d%
\mathbf{r}_{2}}{\left\vert \int S(\mathbf{r};\omega )d\mathbf{r}\right\vert
^{2}}.  \label{2.16}
\end{equation}%
According to its definition the values of the parameter $\zeta (\omega )$
lie in the range 0$\leq \zeta (\omega )\leq 1$.

For narrow bandwidth light the complex degree of coherence can be
approximated by $\gamma (\mathbf{r}_{1},\mathbf{r}_{2};\tau )\approx \gamma (%
\mathbf{r}_{1},\mathbf{r}_{2};0)\exp (-i\bar\omega \tau )$ leading to a
relationship $|\gamma (\mathbf{r}_{1},\mathbf{r}_{2};\tau )|=|\mu (\mathbf{r}%
_{1},\mathbf{r}_{2};\bar\omega )|$, where $\bar\omega $ is the average
frequency. In the same conditions, similar relationships can be obtained for
the spectral density and the average intensity $S(\mathbf{r},\omega
)=\left\langle I(\mathbf{r})\right\rangle \delta (\omega -\bar\omega )$,
where $\delta (\omega )$ is the delta function. As a consequence, for narrow
bandwidth light, the spectral density is equivalent to the average intensity.


\subsection{Coherent-mode representation of correlation functions}

It has been shown \cite{MW95}, that under very general conditions, one can
represent the CSD of a partially coherent, statistically stationary field of
any state of coherence as a sum of independent coherent modes 
\begin{equation}
W(\mathbf{r}_{1},\mathbf{r}_{2};\omega )=\sum_{j}\beta _{j}(\omega
)E_{j}^{\ast }(\mathbf{r}_{1};\omega )E_{j}(\mathbf{r}_{2};\omega ),
\label{2.11}
\end{equation}%
where $\beta _{j}(\omega )$ and $E_{j}(\mathbf{r};\omega )$ are the
eigenvalues and eigenfunctions, that satisfy the Fredholm integral equation
of the second kind%
\begin{equation}
\int W(\mathbf{r}_{1},\mathbf{r}_{2};\omega )E_{j}(\mathbf{r}_{1};\omega )d%
\mathbf{r}_{1}=\beta _{j}(\omega )E_{j}(\mathbf{r}_{2};\omega ).
\label{2.12}
\end{equation}%
The eigenfunctions in (\ref{2.11}-\ref{2.12}) form an orthogonal set.

According to the definition of the spectral density, $S(\mathbf{r},\omega )$, (%
\ref{2.6}) we have in the case of a coherent-mode representation of the
fields%
\begin{equation}
S(\mathbf{r},\omega )=\sum_{j}\beta _{j}(\omega )|E_{j}(\mathbf{r};\omega
)|^{2}.  \label{2.13}
\end{equation}


\subsection{Propagation of the wavefield correlation functions in the free
space}

For our purposes it is especially important to calculate correlation
functions at different distances from the source. Propagation of the
cross-spectral density $W(\mathbf{r}_{1},\mathbf{r}_{2};\omega )$ in the
half space $z>0$ from the source plane at $z=0$ to the plane at distance $z$
in paraxial approximation is determined by the following expression \cite%
{MW95} 
\begin{equation}
W(\mathbf{r}_{1},\mathbf{r}_{2},z;\omega )=\int\limits_{\Sigma
}\int\limits_{\Sigma }W_{S}(\mathbf{s}_{1},\mathbf{s}_{2},0;\omega
)P_{z}^{\ast }(\mathbf{r}_{1}-\mathbf{s}_{1};\omega )P_{z}(\mathbf{r}_{2}-%
\mathbf{s}_{2};\omega )d\mathbf{s}_{1}d\mathbf{s}_{2},  \label{2.9}
\end{equation}%
where $W_{S}(\mathbf{s}_{1},\mathbf{s}_{2},0;\omega )$ is the value of the CSD
at the source plane $z=0$, $P_{z}(\mathbf{r}-\mathbf{s;}\omega )$ is the
Green function (or propagator), and the integration is made in the source
plane. In Eq. (\ref{2.9}) and below coordinates $\mathbf{s}_{1}$ and $\mathbf{%
s}_{2}$ are taken in the plane of the source and coordinates $\mathbf{r}_{1}$
and $\mathbf{r}_{2}$ are taken in the plane at the distance $z$. The
propagator $P_{z}(\mathbf{r}-\mathbf{s;}\omega )$ describes the propagation
of radiation in free space and is defined as

\begin{equation}
P_{z}(\mathbf{r}-\mathbf{s;}\omega )=\frac{1}{i\lambda z}\exp \left[ i\frac{k%
}{2z}\left( \mathbf{r}-\mathbf{s}\right) ^{2}\right] ,  \label{2.10}
\end{equation}%
where $k=2\pi /\lambda $ and $\lambda $ is the wavelength of radiation.

In the case of the coherent-mode representation of the correlation
functions, the values of the CSD can be obtained at different distances by
propagating the individual coherent modes. Propagation of the individual
modes in the half space $z>0$ from the source plane at $z=0$ to the plane at
distance $z$ in the paraxial approximation can be obtained from 
\begin{equation}
E_{j}(\mathbf{r},z;\omega )=\int\limits_{\Sigma }E_{j}^{S}(\mathbf{s}%
;\omega )P_{z}(\mathbf{r}-\mathbf{s;}\omega )d\mathbf{s,}  \label{2.15}
\end{equation}%
where $E_{j}^{S}(\mathbf{s};\omega )$ are the values of the field amplitudes
at the source plane, $z=0$, and $P_{z}(\mathbf{r}-\mathbf{s;}\omega )$ is the
same propagator as in (\ref{2.10}). Due to the statistical independence of
the modes \cite{MW95}, the CSD, after propagating a distance $z$, is given
as a sum of propagated modes $E_{j}(\mathbf{r},z;\omega )$ with the same
eigenvalues $\beta _{j}(\omega )$ defined as in (\ref{2.12}) 
\begin{equation}
W(\mathbf{r}_{1},\mathbf{r}_{2},z;\omega )=\sum_{j}\beta _{j}(\omega
)E_{j}^{\ast }(\mathbf{r}_{1},z;\omega )E_{j}(\mathbf{r}_{2},z;\omega ).
\label{5}
\end{equation}

\section{Coherence properties of 3-rd generation synchrotron sources}


\subsection{Gaussian Schell-model source}

In this section we apply the general theory of the propagation of Gaussian
optical beams from a source of any state of coherence to the case of
synchrotron radiation. We assume that a \textit{real }synchrotron source
(for example an insertion device like an undulator) can be represented by
its \textit{equivalent} model that produces x-ray radiation with statistical
properties similar to a real x-ray source. This equivalent model source will
be positioned in the center of the real undulator source.

We will assume that the x-ray radiation is generated by a planar Gaussian
Schell-model (GSM) source \cite{Footnote2}. Such sources are described by
the following CSD function \cite{MW95} 
\begin{equation}
W_{S}(\mathbf{s}_{1},\mathbf{s}_{2})=\sqrt{S_{S}(\mathbf{s}_{1})}\sqrt{S_{S}(%
\mathbf{s}_{2})}\mu _{S}(\mathbf{s}_{2}-\mathbf{s}_{1}),  \label{3.1}
\end{equation}%
where the spectral density and SDC of the x-ray beam in the source plane are
Gaussian functions\footnote{%
In this equation and below the frequency dependence $\omega $ is omitted.}%
\begin{eqnarray}
S_{S}(\mathbf{s}) &=&S_{0x}S_{0y}\exp \left( -\frac{s_{x}^{2}}{2\sigma
_{Sx}^{2}}-\frac{s_{y}^{2}}{2\sigma _{Sy}^{2}}\right) ,  \notag \\
\mu _{S}(\mathbf{s}_{2}-\mathbf{s}_{1}) &=&\exp \left( -\frac{%
(s_{2x}-s_{1x})^{2}}{2\xi _{Sx}^{2}}-\frac{(s_{2y}-s_{1y})^{2}}{2\xi
_{Sy}^{2}}\right) .  \label{3.2}
\end{eqnarray}%
Here parameters $\sigma _{Sx,y}$ define the rms source size in the $x$- and $%
y$- directions and $\xi _{Sx,y}$ give the coherence length of the source in
the respective directions.

The starting expression for the source cross-spectral density function in
the form of expression (\ref{3.1}), is in fact, very general and is based on
the definition of the SDC (\ref{2.7}). Here, the main approximations are
that the source is modeled as a plane two-dimensional source, that the
spectral density, $S_{S}(\mathbf{s})$, and the SDC, $\mu _{S}(\mathbf{s}_{2}-%
\mathbf{s}_{1})$, (\ref{3.2}) are Gaussian functions, and that the source is
spatially uniform (SDC $\mu _{S}(\mathbf{s}_{2}-\mathbf{s}_{1})$ depends
only on the difference of spatial coordinates $\mathbf{s}_{1}$ and $\mathbf{s%
}_{2}$). The fact that synchrotron radiation sources are typically elongated
in the horizontal direction is specifically introduced in the expression (%
\ref{3.2}) by allowing the source size $\sigma _{Sx,y}$ and coherence
length of the source $\xi _{Sx,y}$ to be different in $x$- and $y$-
direction. What is especially important in this model is the assumption of a
certain degree of coherence of the source expressed by a finite coherence
length of that source $\xi _{Sx,y}$. Only with this finite coherence length
of the source it is possible to get a reasonable description of the
synchrotron radiation with its extremely small divergence.

It can be shown \cite{MW95}, that with a suitable choice of source size, $%
\sigma _{S}$, and coherence length, $\xi _{S}$, a GSM source can generate a
field whose intensity has appreciable values only within a narrow cone of
solid angle. In optics radiation with a narrow angular divergence is called
a beam. This will be a good model for x-ray beams generated by insertion
devices, such as undulators, which produce x-ray beams with a divergence of a
few micro radians. In order to generate the beam, the parameters of a GSM
source have to satisfy the following inequality in each direction%
\begin{equation}
\frac{1}{\delta _{Sx,y}^{2}}\ll \frac{2\pi ^{2}}{\lambda ^{2}},  \label{3.3}
\end{equation}%
where 
\begin{equation}
\frac{1}{\delta _{Sx,y}^{2}}=\frac{1}{(2\sigma _{Sx,y})^{2}}+\frac{1}{\xi
_{Sx,y}^{2}}. \label{3.3c}
\end{equation}%
For x-ray wavelengths of about 0.1 nm we get for the right hand side of this
inequality $2\cdot 10^{9}$ $1/\mu m^{2}$. The smallest size of a source that
is at the moment available at 3-rd generation synchrotron in the vertical
direction is of the order of a few microns. We have not yet estimated the
coherence length $\xi _{S}$ of the source, but as we will see later it is
also of the order of a few microns. From these estimates, it is seen
that the beam condition (\ref{3.3}) is very well satisfied for x-ray
wavelengths and 3-rd generation synchrotron sources. This gives us
confidence using the beam approach to describe the properties of the x-ray
radiation from these sources.

There are two important limits which we can describe as an incoherent or
coherent source. The source will be called \textit{incoherent} if its
coherence length $\xi _{S}$ is much smaller than the source size $\xi
_{S}\ll \sigma _{S}$. From the beam condition (\ref{3.3}) we find for this
source 
\begin{equation*}
\delta _{S}\approx \xi _{S}\gg \frac{1}{\sqrt{2}\pi }\lambda.
\end{equation*}%
This means that to satisfy the beam conditions for a spatially incoherent
source the coherence length has to be small but at the same time larger than
the wavelength $\sigma _{S}\gg $ $\xi _{S}\gg \lambda $. In the opposite
limit of a spatially \textit{coherent} source $\xi _{S}\gg \sigma _{S}$ we
find from the beam condition (\ref{3.3}) 
\begin{equation*}
\delta _{S}\approx 2\sigma _{S}\gg \frac{1}{\sqrt{2}\pi }\lambda .
\end{equation*}%
This means that to satisfy the beam condition for a spatially coherent
source the source size should be larger than the wavelength $\xi _{S}\gg
\sigma _{S}\gg \lambda $.

Integration in (\ref{2.9}) with the CSD $W_{S}(\mathbf{s}_{1},\mathbf{s}%
_{2}) $ (\ref{3.1}, \ref{3.2}) can be done independently for each dimension.
This gives the following expression for the CSD $W(x_{1},x_{2},z)$\footnote{%
Below we present results only in \textit{x}-direction. The same equations
are valid in \textit{y}-direction.} at distance z from the source \cite%
{FS82,FS83} (see also Mandel \& Wolf \cite{MW95}) 
\begin{equation}
W(x_{1},x_{2},z)=\frac{I_{0x}}{\Delta _{x}(z)}e^{i\psi _{x}(z)}\exp \left[ -%
\frac{(x_{1}+x_{2})^{2}}{8\sigma _{Sx}^{2}\Delta _{x}^{2}(z)}\right] \exp %
\left[ -\frac{(x_{2}-x_{1})^{2}}{2\delta _{Sx}^{2}\Delta _{x}^{2}(z)}\right]
.  \label{3.4}
\end{equation}%
Here 
\begin{equation}
\Delta _{x}(z)=\left[ 1+\left( \frac{z}{z_{x}^{eff}}\right) ^{2}\right]
^{1/2}  \label{3.5a}
\end{equation}%
is called an expansion coefficient and 
\begin{equation}
\psi _{x}(z)=\frac{k(x_{2}^{2}-x_{1}^{2})}{2R_{x}(z)},R_{x}(z)=z\left[
1+\left( \frac{z_{x}^{eff}}{z}\right) ^{2}\right]  \label{3.5b}
\end{equation}%
are the phase and the radius of curvature of a Gaussian beam. In Eqs. (\ref%
{3.5a}, \ref{3.5b}) an effective distance $z_{x}^{eff}$ is introduced that
is defined as%
\begin{equation}
z_{x}^{eff}=k\sigma _{Sx}\delta _{Sx}.  \label{3.5d}
\end{equation}%
At that distance the expansion coefficient $\Delta $($z_{x}^{eff}$)=$\sqrt{2}
$. In the limit of a spatially coherent source $\delta _{S}\approx 2\sigma
_{S}$ and an effective distance $z_{x}^{eff}$ coincides with the so-called
Rayleigh length $z_{R}=2k\sigma _{Sx}^{2}$, which is often introduced in the
theory of optical Gaussian beams \cite{Siegman}. According to Eq. (\ref{3.5a}%
) an effective distance $z_{x}^{eff}$ can serve as a measure of the
distances, where nonlinear effects in the propagation of the beams are still
strong. Distances $z\gg z_{x}^{eff}$ can be considered as a far-field limit
where the expansion parameter $\Delta _{x}(z)\rightarrow z/z_{x}^{eff}$ and the
radius $R_{x}(z)\rightarrow z$ change linearly with the distance $z$.

Setting $x_{1}=x_{2}=x$ in Eq. (\ref{3.4}) we obtain for the spectral density%
\begin{equation}
S(x,z)=\frac{I_{0x}}{\Delta _{x}(z)}\exp \left[ -\frac{x^{2}}{2\Sigma
_{x}^{2}(z)}\right] ,  \label{3.4a}
\end{equation}%
where%
\begin{equation}
\Sigma _{x}(z)=\sigma _{Sx}\Delta _{x}(z)=\left[ \sigma _{Sx}^{2}+\theta
_{\Sigma x}^{2}z^{2}\right] ^{1/2}  \label{3.4c}
\end{equation}%
is an rms size of the x-ray beam at a distance $z$ from the source. In Eq. (%
\ref{3.4c}) $\theta _{\Sigma x}$ is the angular divergence of the beam%
\begin{equation}
\theta _{\Sigma x}=\frac{1}{2k\xi _{Sx}}\left[ 4+q_{Sx}^{2}\right] ^{1/2}.
\label{3.6b}
\end{equation}%
Here a ratio of the coherence length to the source size 
\begin{equation}
q_{Sx}=\frac{\xi _{Sx}}{\sigma _{Sx}}  \label{3.6e}
\end{equation}%
is introduced. It can be considered as a measure of the degree of coherence
of the source.

According to its definition (\ref{2.7}) and (\ref{3.4}, \ref{3.4a}), the
spectral degree of coherence at a distance $z$ from the source is given by 
\begin{equation}
\mu (x_{1},x_{2},z)=e^{i\psi _{x}(z)}\exp \left[ -\frac{(x_{2}-x_{1})^{2}}{%
2\Xi _{x}^{2}(z)}\right] ,  \label{3.4b}
\end{equation}%
where 
\begin{equation}
\Xi _{x}(z)=\xi _{Sx}\Delta _{x}(z)=\left[ \xi _{Sx}^{2}+\theta _{\Xi
x}^{2}z^{2}\right] ^{1/2}  \label{3.4d}
\end{equation}%
is the effective coherence length of an x-ray beam at the same distance. In
Eq. (\ref{3.4d}) $\theta _{\Xi x}$ is the angular width of the coherent part
of the beam%
\begin{equation}
\theta _{\Xi x}=\frac{1}{2k\sigma _{Sx}}\left[ 4+q_{Sx}^{2}\right] ^{1/2}.
\label{3.6d}
\end{equation}

For the \textit{incoherent} source ($q_{Sx}\ll 1$) we have from (\ref{3.6b}, %
\ref{3.6d})%
\begin{equation}
\theta _{\Sigma x}=\frac{1}{k\xi _{Sx}},\theta _{\Xi x}=\frac{1}{k\sigma
_{Sx}}.  \label{3.7a}
\end{equation}%
It is seen immediately that in this limit equation (\ref{3.4d}) predicts the
same values for the coherence length $\Xi _{x}(z)$ at large distances $z$ as
given by the van Zittert-Cernike theorem. At the same time expression (\ref%
{3.7a}) gives an estimate for the divergence of the beam from an incoherent
source, which is determined by the coherence length $\xi _{Sx}$ of the
source. So, directly from (\ref{3.7a}), we have an estimate of the coherence
length of the incoherent source 
\begin{equation}
\xi _{Sx}=\frac{\lambda }{2\pi \theta _{\Sigma x}}.  \label{3.8}
\end{equation}

In another limit of a \textit{coherent} source, when parameter $q_{Sx}\gg 1$%
, we obtain from (\ref{3.6b}, \ref{3.6d})%
\begin{equation}
\theta _{\Sigma x}=\frac{1}{2k\sigma _{Sx}},\theta _{\Xi x}=\frac{1}{%
2k\sigma _{Sx}}q_{Sx}.  \label{3.9a}
\end{equation}
In this coherent limit the angular width of the coherent part of the beam
exceeds the angular divergence of the beam, which is determined now only by
the size of the source, and we are approaching here the limit of a so-called
diffraction limited source.

In the frame of the GSM, the coherence length of the source of any state of
coherence can be expressed conveniently through its emittance $\varepsilon
_{Sx}=\sigma _{Sx}\sigma _{Sx}^{\prime }$, where $\sigma _{Sx}^{\prime }$ is
the rms of the angular divergence of the source. It can be obtained by
inverting the full expression of the angular divergence of the beam (\ref%
{3.6b})

\begin{equation}
\xi _{Sx}=\frac{2\sigma _{Sx}}{\sqrt{4k^{2}\varepsilon _{Sx}^{2}-1}}
\label{3.11}
\end{equation}%
and the substitution of the angular divergence $\theta _{\Sigma x}$ by $%
\sigma _{Sx}^{\prime }$.

One important property of the beams generated by the GSM sources is that at
any distance from the source the ratio of the coherence length $\Xi _{x}(z)$
to the beam size $\Sigma _{x}(z)$ is a constant value and is equal to the
same ratio at the source. From Eqs. (\ref{3.4c}, \ref{3.4d}) we have for the
parameter $q_{x}$ 
\begin{equation}
q_{x}=\frac{\xi _{Sx}}{\sigma _{Sx}}=\frac{\Xi _{x}(z)}{\Sigma _{x}(z)}.
\label{3.13}
\end{equation}%
The degree of transverse coherence $\zeta _{x}$ introduced in Eq. (\ref{2.16}%
) can be directly calculated for the GSM source and related to the values of 
$q_{x}$ by the following expression%
\begin{equation}
\zeta _{x}=\frac{q_{x}}{\sqrt{q_{x}^{2}+4}}.  \label{3.13a}
\end{equation}%
According to the relationship (\ref{3.13}) the values of the degree of
transverse coherence $\zeta _{x}$ for the GSM are preserved for any distance 
$z$ from the source.

The emittance of a GSM source $\varepsilon _{Sx}$ can be expressed through the
degree of the transverse coherence $\zeta _{x}$. From Eqs. (\ref{3.6b}) and (%
\ref{3.13a}) we have for the emittance of a GSM source 
\begin{equation}
\varepsilon _{Sx}=\frac{1}{2k\zeta _{x}}.  \label{3.13c}
\end{equation}%
Taking into account that for a source of any degree of coherence the values
of $\zeta _{x}$ lie in the range $0\leq \zeta _{x}\leq 1$ the values of the
emittance should satisfy an inequality $\varepsilon _{Sx}\geq 1/2k=\lambda
/4\pi $. For a fully coherent source $\zeta _{x}\rightarrow 1$ and the
emittance $\varepsilon _{Sx}^{coh}=\lambda /4\pi $. This value can be
considered as the emittance of a diffraction limited source. For an
incoherent source $\zeta _{x}\rightarrow 0$ and according to Eq. (\ref{3.13c}%
) $\varepsilon _{Sx}\gg\lambda /4\pi $.


\subsection{Transverse coherence properties of the PETRA\ III source}

Our previous analysis can be effectively used to estimate the coherence
properties of the beams produced by 3-rd generation x-ray sources if source
parameters (source size and divergence) are known. We will make this
calculation for the high brilliance synchrotron source PETRA III that is
presently under construction at DESY. This storage ring is planned to
produce $\varepsilon _{x}=1$ nm emittance beams in the horizontal direction
and, due to 1\% coupling, the emittance in the vertical direction will be two
orders of magnitude lower.
\begin{table}[tbp]
\centering           
\caption{Parameters of the high brilliance synchrotron radiation
source {\ PETRA III} for a 5 m undulator \cite{PETRA} (energy $E$=12 keV,
distance from the source $z$=60 m) }
\begin{tabular}{|r|c|c|c|c|}
	\multicolumn{2}{c}{ }\\
\hline
& \multicolumn{2}{|c|}{\textbf{High-}${\beta}$} & \multicolumn{2}{|c|}{%
\textbf{Low-}${\beta}$} \\ \cline{2-5}
& $x$ & $y$ & $x$ & $y$ \\ \hline
Source size $\sigma_S,~[\mu$m] & 141 & 5.5 & 36 & 6 \\ 
Source divergence $\sigma_S^{\prime},~[\mu$rad] & 7.7 & 3.8 & 28 & 3.7 \\ 
Transverse coherence length &  &  &  &  \\ 
at the source $\xi_S,~[\mu$m] & 2.07 & 4.53 & 0.57 & 4.65 \\ 
Degree of coherence $q$ & 0.015 & 0.82 & 0.016 & 0.77 \\ 
Degree of transverse coherence $\zeta$ & 0.008 & 0.38 & 0.008 & 0.36 \\ 
Effective length $z^{eff}$, [m] & 18.33 & 1.48 & 1.29 & 1.63 \\ 
Beam size at distance $z~\Sigma(z)$, [mm] & 0.48 & 0.23 & 1.68 & 0.22 \\ 
Transverse coherence length &  &  &  &  \\ 
at distance $z~\Xi(z),~[\mu$m] & 7.08 & 187.8 & 26.5 & 170.5 \\ \hline
\end{tabular}%
\label{table1}
\end{table}

Source parameters for a 5 m long undulator and a photon energy of 12 keV are
summarized in Table \ref{table1}. Two cases of high-$\beta $ and low-$\beta $
operation are considered. The values of the coherence length of the source
calculated according to Eq. (\ref{3.11}) vary from 0.6 to 2 microns in the
horizontal direction and are about 5 microns in the vertical. We can
estimate the values of the parameter $q_{S}$ (\ref{3.13}) and the degree of
transverse coherence $\zeta _{S}$ of that source. Using tabulated values of
the source size we get for the horizontal direction $q_{Sx}\sim 0.02$ and
for the vertical $q_{Sy}\sim 0.8$. For the degree of transverse coherence $%
\zeta _{S}$ (\ref{3.13a}) we have in the horizontal direction $\zeta
_{Sx}\sim 0.01$ and in the vertical $\zeta _{Sy}\sim 0.4$. These estimates
immediately show that in the horizontal direction PETRA III source is a
rather incoherent source with the degree of coherence about 1\%, however in
the vertical direction the coherence length of the source is about the size
of the source itself, and it can be considered as a rather coherent source
with a degree of coherence about 40\%. Substituting these numbers into (\ref%
{3.4c}, \ref{3.4d}) we can obtain the values of the intensity distribution
and the transverse coherence length at any distance downstream from the
source. These values are listed in Table \ref{table1} for a distance $z=60$ $%
m$, where the first experimental hutches are planned. We see that for this
distance the coherence length is varying from $7$ $\mu m$ to $25$ $\mu m$ in
the horizontal direction and is in the range from 170 $\mu $m$\ $to 190 $\mu 
$m in the vertical one. This defines the coherence area across the
beam within which one can plan experiments with coherent beams.
\begin{figure}[tbp]
	\centering
	\includegraphics[width=0.9\textwidth]{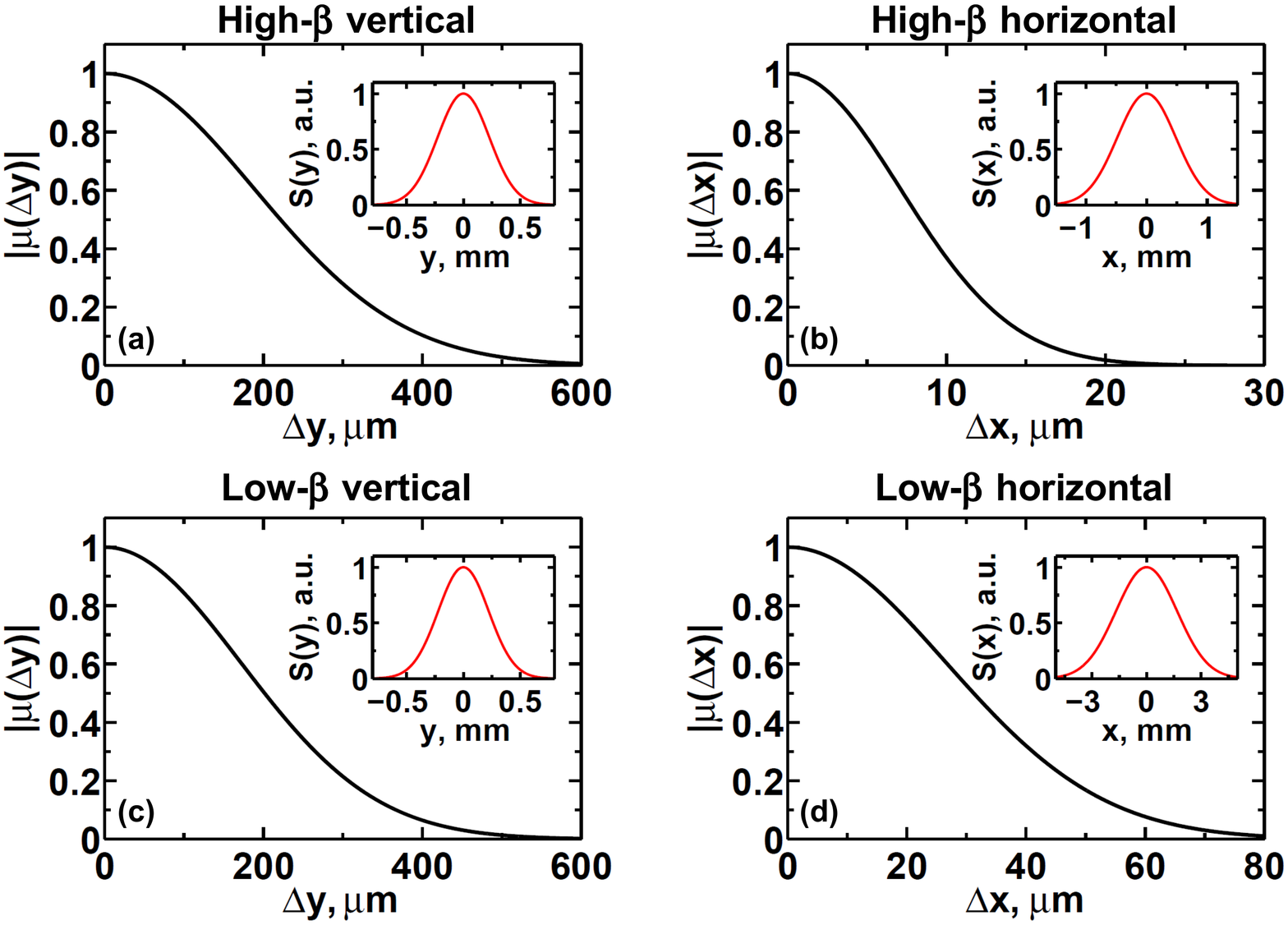} %
	\caption{The absolute value of the spectral degree of coherence $%
\left\vert \mu (\Delta x)\right\vert $ as a function of separation of two
points across the beam at a distance of 60 m downstream from the source. The
spectral density $S(x)$ as a function of the position across the beam
calculated at the same distance from the source is shown in the insets. The
rms values of the beam size $\Sigma _{x,y}(z)$ and transverse coherence
length $\Xi _{x,y}(z)$ at that distance were taken from Table \ref%
{table1}. (a, b) High-$\beta $ vertical and horizontal sections of the beam.
(c, d) Low-$\beta $ vertical and horizontal sections of the beam.
 }
\end{figure}

The absolute value of the SDC, $\left\vert \mu (\Delta x)\right\vert $, (\ref%
{3.4b}) as a function of the separation of two points across the beam and
the spectral density, $S(x)$, (\ref{3.4a}) as a function of the position
across the beam at a distance 60 m downstream from the source are
presented in Fig. 2. The rms values of the beam size $\Sigma _{x,y}(z)$ and
transverse coherence length $\Xi _{x,y}(z)$ were taken from Table \ref%
{table1}. It can be seen in Figs. 2 (a,c) that the properties of the beam in
the vertical direction are very similar for both the high-$\beta $ and low-$%
\beta $ operation of the PETRA III source. The FWHM of the beam is about 500 
$\mu $m in both cases. For separations of up to 100 $\mu $m the beam is
highly coherent (with the degree of coherence higher than 80\%). In the
horizontal direction (Figs. 2 (b,d)) the situation is quite different. The
FWHM of the beam for high-$\beta $ operation is about one millimeter and for
low-$\beta $ operation the beam is quite divergent and its FWHM is about
three millimeters. It is well seen in Figs. 2 (b,d) that the beam is rather
incoherent in the horizontal direction. The degree of coherence is higher
than 80\% for separations of up to 15 $\mu $m for a low-$\beta $ operation
and 5 $\mu $m for a high-$\beta $ operation at this distance.
\begin{figure}[tb]
	\centering
	\includegraphics[width=0.9\textwidth]{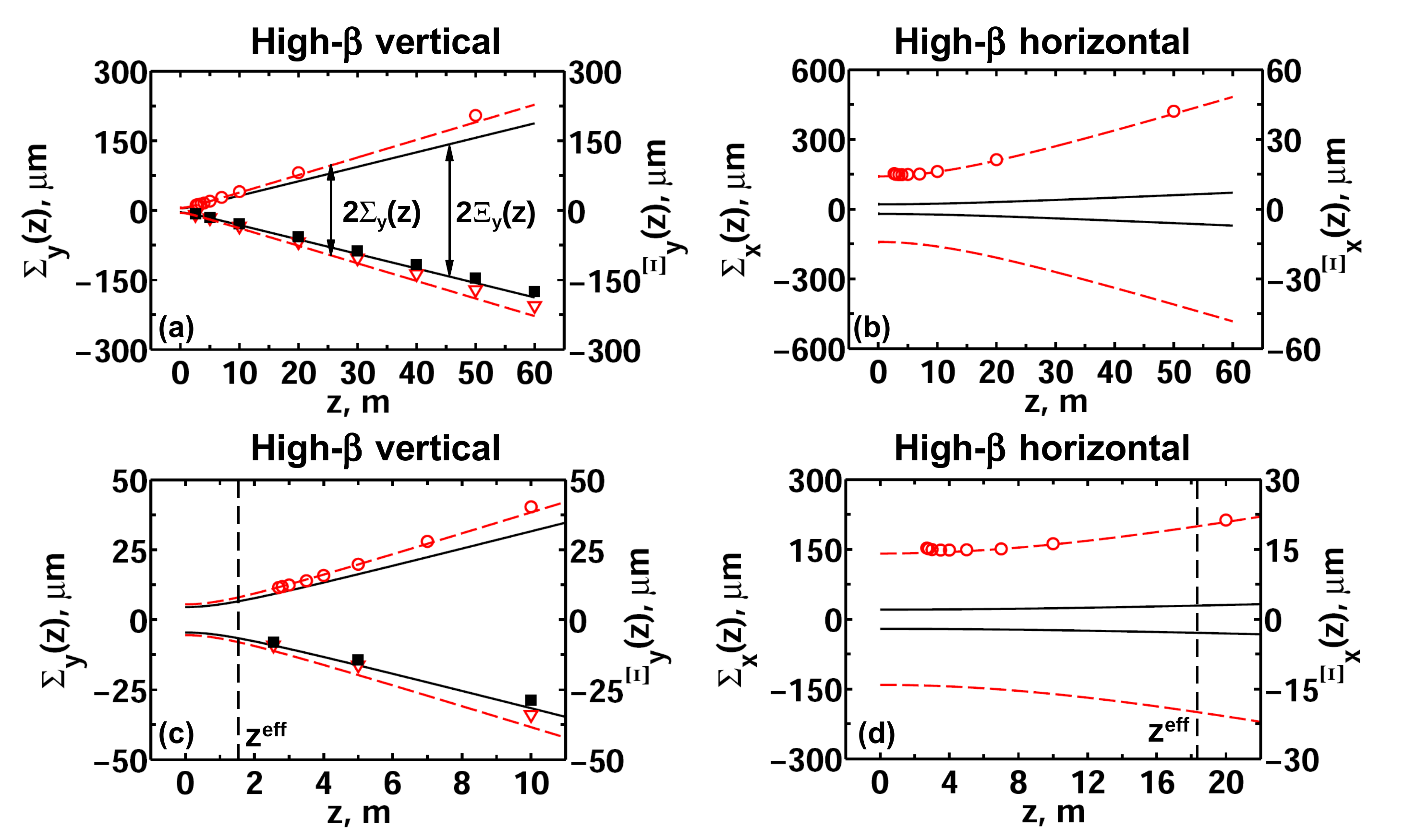} %
	\caption{The beam size $\Sigma _{x,y}(z)$ and the transverse
coherence length $\Xi _{x,y}(z)$ at different distances $z$ from the source
for a high-$\beta $ section of the PETRA III storage ring. Parameters of the
source are taken from Table \ref{table1}. In all figures the dashed (red) line
is the beam size $\Sigma _{x,y}(z)$ and the solid (black) line is the transverse
coherence length $\Xi _{x,y}(z)$. Open circles correspond to calculations
performed by the ESRF simulation code SRW \cite{SRW}, open triangles are the
beam size and squares are the transverse coherence length obtained from the
analytical results of Ref. \cite{GSS08}. (a, c) Vertical direction of the
beam. (b, d) Horizontal direction of the beam. The vertical dashed line in (c)
and (d) correspond to an effective distance $z^{eff}$. Note, different range
for the coherence length comparing to that of the beam size in (b,d). }
\end{figure}

Calculations of the beam size, $\Sigma _{x,y}(z)$, and the transverse
coherence length, $\Xi _{x,y}(z)$, at different distances $z$ from the source
are made according to Eqs. (\ref{3.4c}, \ref{3.4d}). They are presented for
high-$\beta $ operation in Fig. 3 and for low-$\beta $ operation in Fig. 4.
These calculations show that in the vertical direction the properties of the
beam along the beamline for both high-$\beta $ and low-$\beta $ operations
of the PETRA III source are quite similar. The rms values of the coherence
length, $\Xi _{x,y}(z)$, (black, solid line) are slightly smaller than the rms values of the beam
size, $\Sigma _{x,y}(z)$, (red, dash line) along the beamline. At distances larger than $%
z_{y}^{eff}\simeq 1.5$ m in the vertical direction all $z$-dependencies can
be considered to be linear. It means that for all practical cases all
parameters scale linearly with the distance $z$. In the horizontal direction
the situation is quite different. Firstly as expected and clearly seen in
Figs. 3 and 4, the beam is quite incoherent in this direction. It is
also three times more divergent in the far-field for low-$\beta $ operation
(Fig. 4 (b)). It is interesting to note that the linear $z$-dependence of
parameters $\Sigma _{x}(z)$ and $\Xi _{x}(z)$ for high-$\beta $ operation
starts from further distances from the source. An effective distance $%
z_{y}^{eff}$ is about 20 m in this case.

\begin{figure}[tbp]
	\centering
	\includegraphics[width=0.9\textwidth]{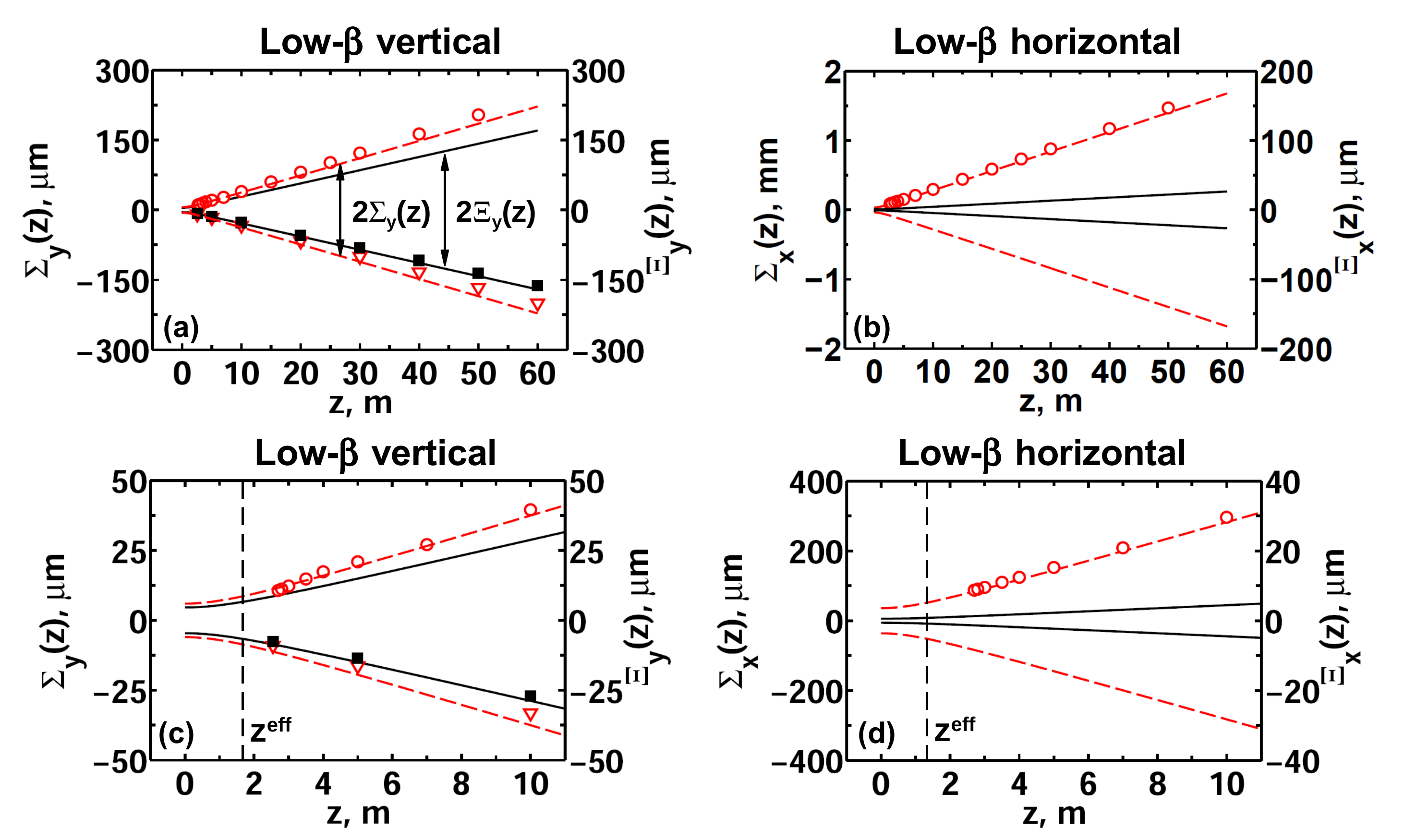} %
	\caption{The same as in Fig. 3 for a low-$\beta $ section of the
PETRA III storage ring.}
\end{figure}

We compared the results obtained by our approach with the results of
different calculations performed for the PETRA III five meter undulator
source by the ESRF simulation code SRW \cite{SRW} as well as by the
analytical results obtained in Ref. \cite{GSS08} (see Appendix for details).
Our calculations show that the divergence of the beam both in the vertical
and in the horizontal directions is well described by our model as compared
with the SRW calculations (see Fig. 3 and Fig. 4). The comparison with the
results of Ref. \cite{GSS08} was performed only in the vertical direction as
in the horizontal direction the source can be described as a
quasi-homogeneous source (coherence length of such a source is much smaller
than the size of the source ($\xi _{Sx}\ll\sigma _{Sx}$)). In this limit the
analytical results of Ref. \cite{GSS08} completely coincide with our
description of the source in the frame of the GSM. However, in the vertical
direction a more careful analysis is required. Using the approach of Ref. 
\cite{GSS08} we calculated the SDC and spectral density for a five meter
undulator of the PETRA III source in the vertical direction at different
distances $z$ from the source. From these calculations we obtained the rms
values of the source size $\Sigma _{x}(z)$ and the coherence length $\Xi
_{x}(z)$ at different distances from the source and compared them with the
results obtained from the GSM (see Figs. 3 (a,c) and Figs 4 (a,c)). This
comparison shows very good agreement between two approaches for these
energies. However, the analytical approach of Ref. \cite{GSS08} gives
slightly lower values of the beam size and the coherence length at larger
distances. The lower values of the beam size predicted by Ref. 
\cite{GSS08} can be attributed to effects of a final energy spread of
electrons in the bunch that were neglected in calculations. From this
comparison, we see that an approach based on the GSM and
simulations performed by different methods at a wavelength of 0.1 nm give
similar results\footnote{
We performed a similar analysis for different energies and came to the
conclusion that for PETRA III source parameters the GSM model can be safely used
at energies higher than 6 keV (see Appendix for details).}.

\section{Coherence properties of x-ray free-electron lasers}


\subsection{Coherent-mode decomposition for the GSM source}

We apply a general approach of coherent-mode decomposition, described in
section two, for the analysis of the correlation properties of wavefields
originating from XFEL sources. We substitute a real XFEL source by an
equivalent planar GSM source (\ref{3.1}, \ref{3.2}). Coherent modes and
eigenvalues obtained as a solution of the Fredholm integral equation (\ref%
{2.12}) for such a GSM source are well known \cite{G80,SW82} and can be
decomposed for each transverse direction $%
E_{l,m}(s_{x},s_{y})=E_{l}(s_{x})E_{m}(s_{y})$ and $\beta _{l,m}=\beta
_{l}\beta _{m}$. The eigenvalues $\beta _{j}$ for the GSM source have a
power law dependence and the eigenfunctions $E_{j}(s_{x})$ at such a source
are described by the Gaussian Hermite-modes\footnote{%
Due to the symmetry of the Gaussian-Schell model we consider below only one
transverse direction.} \cite{G83} 
\begin{eqnarray}
\beta _{j}/\beta _{0} &=&\kappa ^{j},  \label{4.1a} \\
E_{j}(s_{x}) &=&\frac{k^{1/4}}{\left( \pi z_{x}^{eff}\right) ^{1/4}}\frac{1}{%
\left( 2^{j}j!\right) ^{1/2}}H_{j}\left( \sqrt{\frac{k}{z_{x}^{eff}}}%
s_{x}\right) \exp \left[ -\left( \frac{k}{2z_{x}^{eff}}\right) s_{x}^{2}%
\right] ,  \label{4.1b}
\end{eqnarray}%
where $H_{j}(x)$ are the Hermite polynomials of order $j$, $\beta _{0}=\sqrt{%
8\pi }S_{0x}\sigma _{Sx}\delta _{Sx}/$ $(2\sigma _{Sx}+\delta _{Sx})$, and $%
\kappa =(2\sigma _{Sx}-\delta _{Sx})/(2\sigma _{Sx}+\delta _{Sx})$. The
parameters $\sigma _{Sx}$, $\delta _{Sx}$ and $z_{x}^{eff}$ have the same
meaning as in Eqs. (\ref{3.2}, \ref{3.3c}, \ref{3.5d}).

Equation (\ref{4.1a}) for the eigenvalues of the GSM source gives, in fact,
the relative weights with which the different modes contribute to the CSD of
the source. It can be also expressed through the values of the parameter $%
q_{x}$ (\ref{3.13}) \cite{MW95}, or through the values of the degree of
transverse coherence $\zeta _{x}$ (\ref{3.13a}) 
\begin{equation}
\frac{\beta _{j}}{\beta _{0}}=\left[ \frac{1}{\left( q_{x}^{2}/2\right)
+1+q_{x}\left[ (q_{x}/2)^{2}+1\right] ^{1/2}}\right] ^{j}=\left( \frac{%
1-\zeta _{x}}{1+\zeta _{x}}\right) ^{j}.  \label{4.2}
\end{equation}%
For a spatially coherent source ($\xi _{Sx}\gg \sigma _{Sx}$) we have from
Eq. (\ref{4.2})%
\begin{equation}
\frac{\beta _{j}}{\beta _{0}}\approx q_{x}^{-2j}.  \label{4.3a}
\end{equation}%
According to this equation $\beta _{j}\ll \beta _{0}$ for all $j\neq 0$ that
means that in the coherent limit the source can be well characterized by its
lowest mode. In the opposite limit of an incoherent source ($\xi _{Sx}\ll
\sigma _{Sx}$) we have from Eq. (\ref{4.2})%
\begin{equation}
\frac{\beta _{j}}{\beta _{0}}\approx 1-jq_{x}.  \label{4.3b}
\end{equation}%
According to this equation many modes are necessary for a sufficient
description of the source.

Correlation properties of the fields in the coordinate-frequency domain at
any distance $z$ from the source can be calculated with the help of
expression (\ref{5}) by propagating individual modes $E_{j}(x,z)$. In the
case of the GSM source the propagated modes $E_{j}(x,z)$ at a distance $z$
from the source are described by the following expression \cite{G83} 
\begin{eqnarray}
E_{j}(x,z) &=&\frac{k^{1/4}}{\left( \pi z_{x}^{eff}\Delta _{x}^{2}(z)\right)
^{1/4}}\frac{1}{\left( 2^{j}j!\right) ^{1/2}}H_{j}\left[ \sqrt{\frac{k}{%
z_{x}^{eff}}}\left( \frac{x}{\Delta _{x}(z)}\right) \right] \times  \notag \\
&&\times \exp \left[ -\frac{k}{2z_{x}^{eff}}\left( \frac{x}{\Delta _{x}(z)}%
\right) ^{2}\right] \times  \notag \\
&&\times \exp \left\{ i[kz-(j+1)\phi _{x}(z)]+\frac{ikx^{2}}{2R_{x}(z)}%
\right\} ,  \label{4.4}
\end{eqnarray}%
where $\phi _{x}(z)=\arctan \left( z/z_{x}^{eff}\right) $. Parameters $\Delta _{x}(z)$
and $R_{x}(z)$ have the same meaning as in (\ref{3.5a}, \ref{3.5b}). For $%
j=0 $ these modes coincide with an expression for a monochromatic Gaussian
beam propagating from a Gaussian source.


\subsection{Transverse coherence properties of the European XFEL source}
\begin{table}[!tb]
\centering        
\caption{Parameters of the SASE1 undulator of the European XFEL \cite%
{XFEL}
}
\begin{tabular}{|r|c|}
	\multicolumn{2}{c}{ }\\
\hline
& \textbf{SASE1} \\ 
& \textbf{undulator} \\ \hline
Wavelength $\lambda$, [nm] & 0.1 \\ 
Source size $\sigma_S,~$[$\mu$m] & 29.7 \\ 
Source divergence $\sigma_S^{\prime},~$[$\mu$rad] & 0.43 \\ 
Transverse coherence length &  \\ 
at the source $\xi_S,~$[$\mu$m] & 48.3 \\ 
Degree of coherence $q$ & 1.63 \\ 
Degree of transverse coherence $\zeta$ & 0.63 \\ 
Effective length $z^{eff}$, [m] & 70 \\ \hline
\end{tabular}%
\label{table2}
\end{table}
We used this approach to make a realistic and simple estimate of the
coherence properties of the upcoming XFEL sources. For detailed calculations
we took parameters of the SASE1 undulator at the European XFEL reported in 
\cite{XFEL} (see also Ref. \cite{SSY08}) and summarized in Table \ref{table2}.
Simulations were made
for a GSM source (\ref{3.1}, \ref{3.2}) with an rms source size $\sigma
_{S}=29.7~\mu $m and a transverse coherence length at the source of $\xi
_{S}=48.3~\mu $m. The latter parameter was obtained from Eq. (\ref{3.11})
using the values of the source size and angular divergence listed in Table %
\ref{table2}. With these parameters the CSD, $W(x_{1},x_{2};z)$, (\ref{5}) was
calculated at a distance of 500 m from the source with the eigenvalues $%
\beta _{j}$ and eigenfunctions $E_{j}(x,z)$ evaluated from Eqs. (\ref{4.1a}, %
\ref{4.4}). A distance of 500 m was considered because at this distance the
first optical elements of the European XFEL are planned.\textit{\ }In Fig. 5
the results of these calculations are presented. An analysis of the results
shows that for the parameters of the SASE1 undulator at XFEL a small number
of transverse modes contribute to the total field (Fig. 5 (c)). Parameter $%
\kappa =0.22$ in these conditions, which means that the contribution of the
first mode is about 20\% of the fundamental and the contribution of the
fourth mode is below one per cent of the fundamental $\beta _{4}/\beta
_{0}=\kappa ^{4}=2.3\times 10^{-3}$. Finally, five modes (including the
fundamental) were used in (\ref{5}) for calculations of the CSD $%
W(x_{1},x_{2};z)$ (Fig. 5 (a)). From the obtained values of the CSD, the
modulus of the SDC, $|\mu (x_{1},x_{2},z)|$, (Figs. 5 (b,d)) and the spectral
density, $S(x)$, at that distance were evaluated. As our source is described
as a Gaussian source these functions are Gaussian as well. At a distance $%
z=500$ m from the source we obtained a coherence length $\Xi (z)=348~\mu $m
and a beam size $\Sigma (z)=214~\mu $m.
\begin{figure}[!tb]
	\centering
	\includegraphics[width=0.9\textwidth]{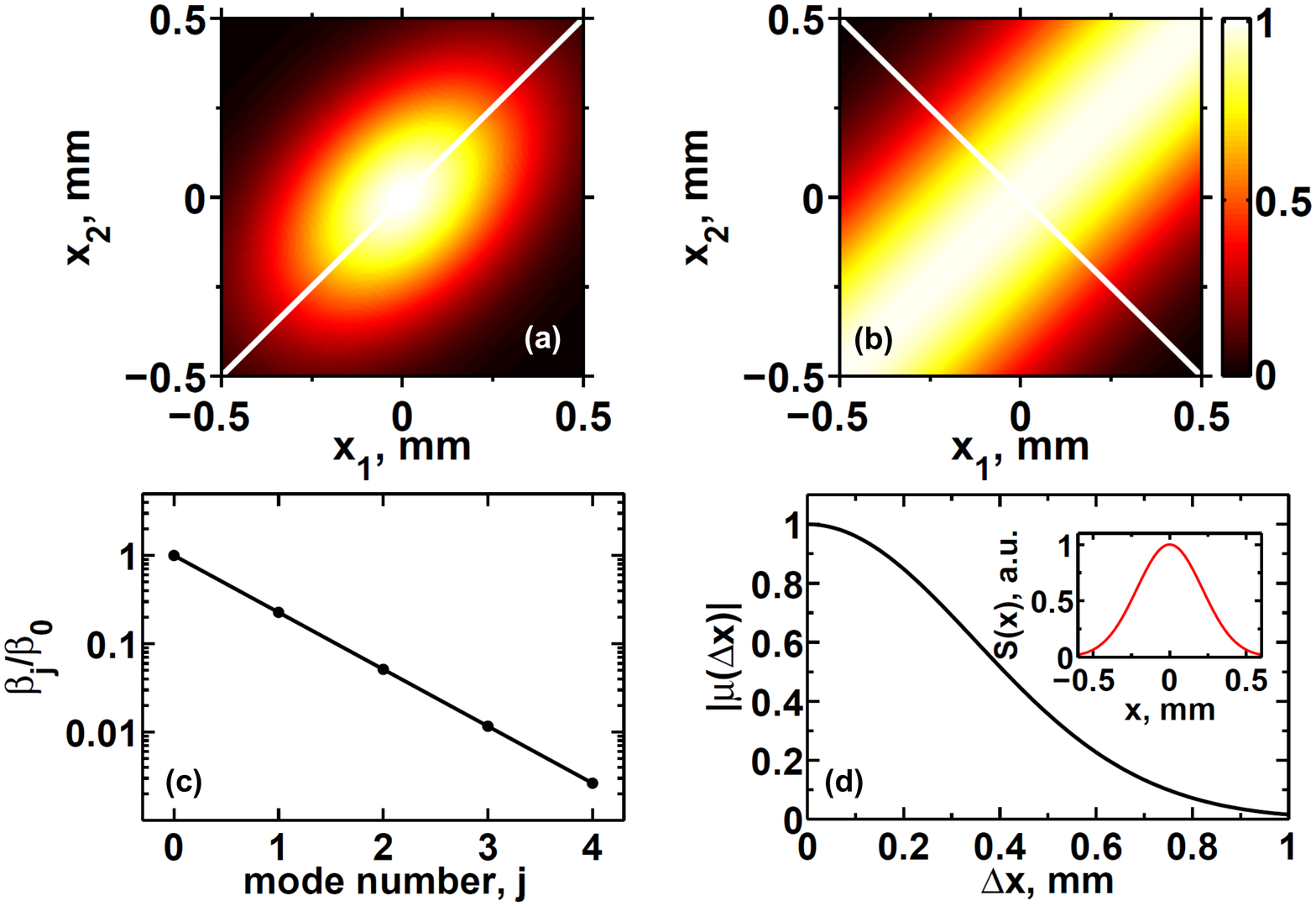} %
	\caption{ Calculations of the coherence properties of the SASE1 undulator
at the European XFEL (see Table \ref{table2}) 500 m downstream from the source in the frame of
a GSM source.
(a) The absolute value of the cross-spectral density $%
\left\vert W(x_{1},x_{2})\right\vert $. (b) The absolute value of the
spectral degree of coherence $\left\vert \mu (x_{1},x_{2})\right\vert $. (c)
The ratio $\beta _{j}/\beta _{0}$ of the eigenvalue $\beta _{j}$ to the
lowest order eigenvalue $\beta _{0}$ as a function of mode number $j$. (d)
The absolute value of the spectral degree of transverse coherence $|\mu
(\Delta x)|$ taken along the white line in (b). In the inset spectral
density $S(x)$ is shown that is taken along the white line in (a). }
\end{figure}

Analysis of Fig. 5 (d) shows that our model source, though being highly
coherent, can not be described as a fully coherent source. The problem lies
in the contribution of the higher modes to the fundamental. This is
illustrated in more detail in Fig. 6 where the spectral degree of coherence,
$|\mu (\Delta x)|$, is calculated with a different number of contributing
modes at separation distances of up to 1 mm where the spectral density $S(x)$
is significant. It is readily seen from this figure that only in
the case of a single mode contribution will an XFEL beam be fully coherent
(Fig. 6 (a)). As soon as the first transverse mode contributes to the
fundamental, the SDC, $|\mu (\Delta x)|$, drops quickly and reaches zero at a
separation distance of $\Delta x\approx 700$ $\mu $m (Fig. 6 (b)). It again
increases up for higher separation distances and reaches the value $|\mu
(\Delta x)|=0.3$ at $\Delta x\approx 1$ mm. 
\begin{figure}[tb]
	\centering
	\includegraphics[width=0.9\textwidth]{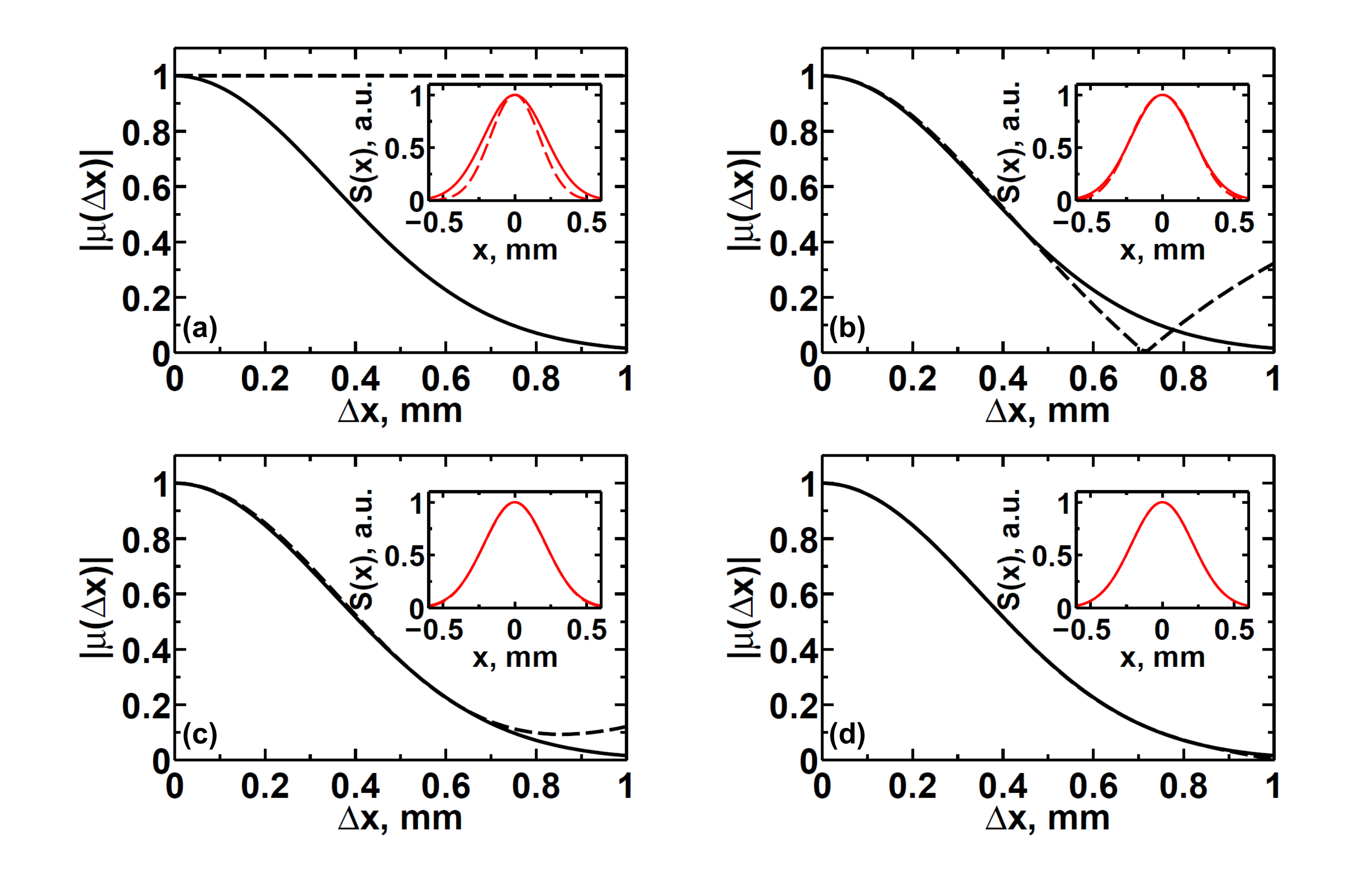} %
	\caption{Contribution of the higher transverse modes to the absolute
value of the spectral degree of coherence $|\mu (\Delta x)|$. The same for
the spectral density $S(x)$ is shown in the insets. (a) Fundamental mode
contribution, (b) fundamental plus first mode contribution, (c) fundamental
plus two modes contribution, (d) fundamental plus three modes contribution.
In all figures the dashed line corresponds to an actual number of modes
contributing to $|\mu (\Delta x)|$ and $S(x)$. In all figures solid line
corresponds to a\ full calculation of $|\mu (\Delta x)|$ and $S(x)$ with the
five modes. Calculations were made for the same parameters as in Fig. 5. }
\end{figure}
This increase in the correlation
function is due to the fact that at these distances the contribution of the
lowest mode (fundamental in this particular case) is negligible and the
correlation properties are determined again by a single mode (the first in
this case). This effect is demonstrated in Fig. 7, where the contribution of
different modes to the spectral density is presented. In this particular
case the spectral density can be well described by three modes.

\begin{figure}
	\centering
	\includegraphics[width=0.6\textwidth]{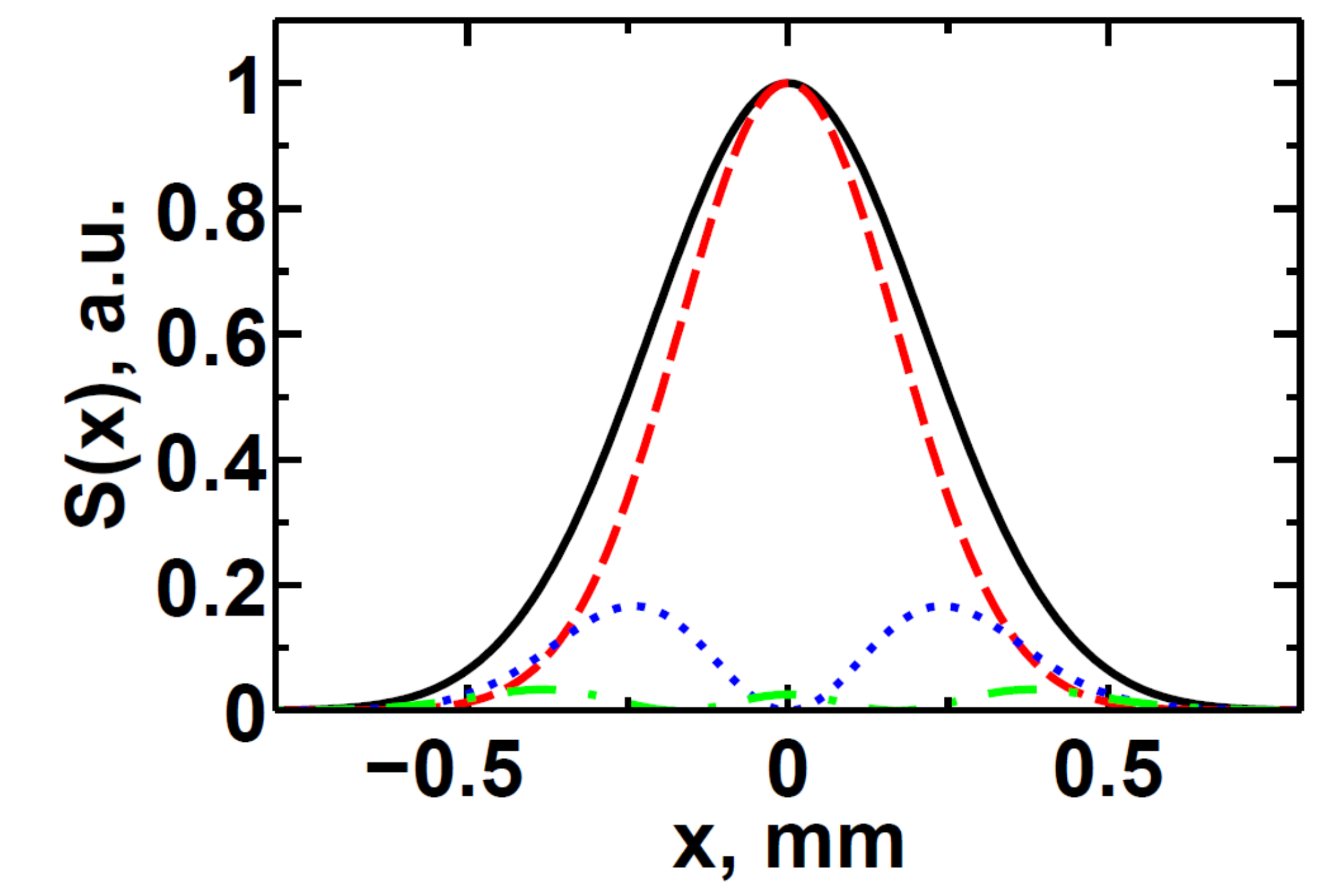} %
	\caption{Contribution of the higher transverse modes to the spectral
density $S(x)$. The solid (black) line corresponds to a full calculation of $S(x)$
with the five modes. The dashed (red) line is the fundamental mode contribution. The
dotted (blue) line is the first mode contribution and dash dotted (green)
line is the second mode contribution. Calculations were made for the same
parameters as in Fig. 5.}
\end{figure}
The values of the beam size $\Sigma (z)$ and the transverse coherence length 
$\Xi (z)$ at different distances $z$ from our GSM source are presented in
Fig. 8. Calculations were performed using a coherent-mode decomposition (\ref%
{5}) of the CSD $W(x_{1},x_{2};z)$ at different distances from the GSM
source. It can be seen from Fig. 8
that contrary to the analysis performed for a synchrotron source, here, in
the case of the European XFEL, the values of the transverse coherence length 
$\Xi (z)$ are higher than the values of the beam size $\Sigma (z)$ at all
distances from the source downstream. An effective distance $z^{eff}$ (\ref%
{3.5d}) is about 70 m in this case, which means that for distances $z\gg 70$
m all $z$-dependencies of parameters, such as the coherence length and the
beam size, can be considered to be linear.

\begin{figure}[tbp]
	\centering
	~~~~~~~~~~~\includegraphics[width=0.71\textwidth]{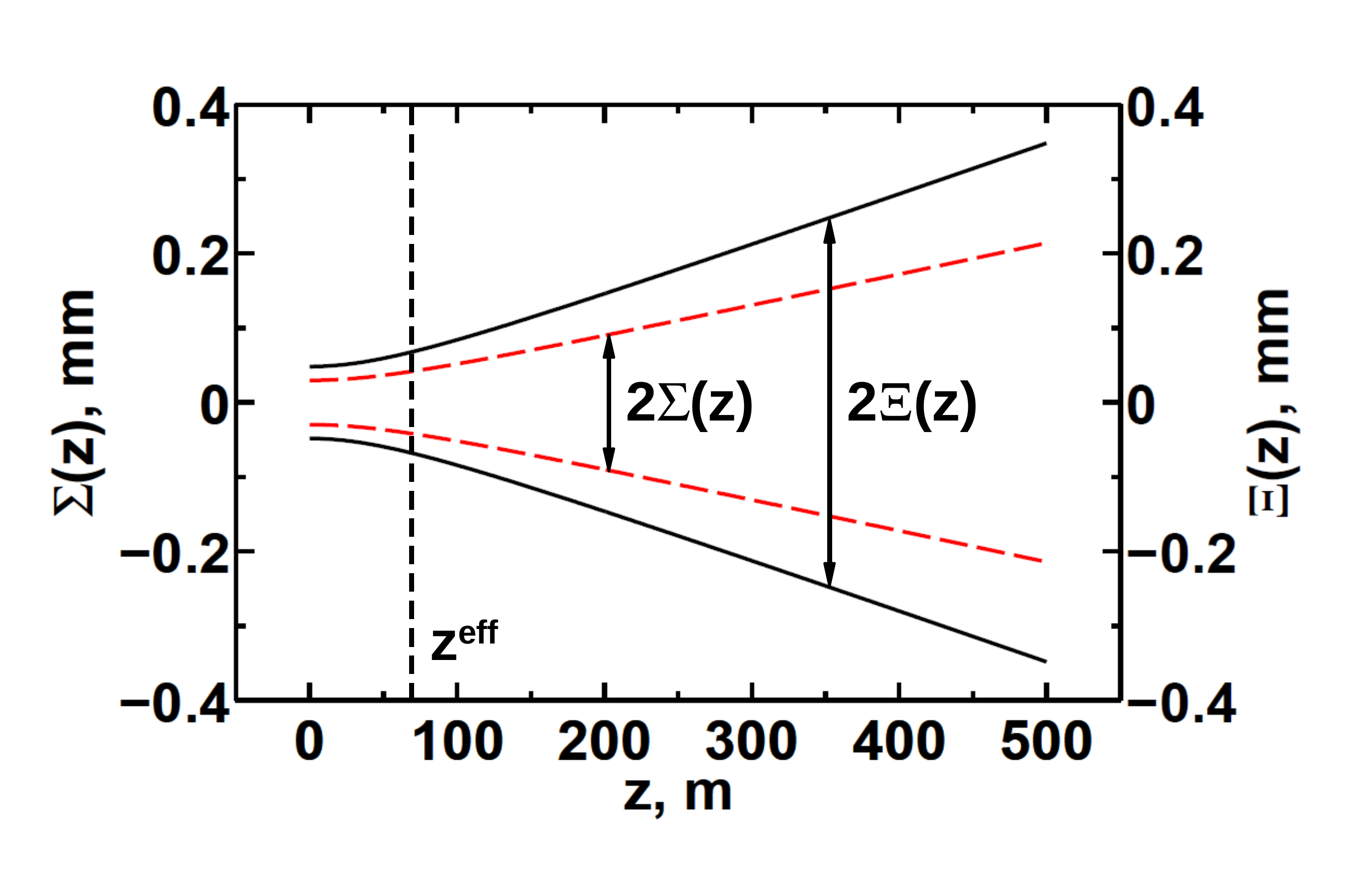} %
	\caption{The beam size $\Sigma (z)$ (dashed line) and the transverse
coherence length $\Xi (z)$ (solid line) at different distances $z$ from the
SASE1 undulator of the European XFEL source. Parameters of the source are
the same as in Fig. 5. The vertical dashed line correspond to an effective
distance $z^{eff}$. }
\end{figure}

Using the previously introduced values of the degree of the transverse
coherence $\zeta $ ( \ref{3.13a}) and the parameters obtained for an
equivalent GSM source (see Table \ref{table2}) we find that $\zeta =0.63$
for that source. This means we can expect a transverse coherence of about
60\% at the European XFEL. This number is in good agreement with the value $%
\zeta =0.65$ obtained by the ensemble average of the wavefields produced by
the SASE1 undulator of the European XFEL calculated by the code FAST using
the actual number of electrons in the beam \cite{SSY08}. This good agreement
obtained for the value of the degree of the transverse coherence $\zeta $ by
different approaches gives good fidelity for the analysis proposed in this
work based on the results of statistical optics and a simple
characterization of the source with a GSM.

Here, for a sufficient description of the transverse coherence properties of
FELs, we used a coherent-mode decomposition approach. In principle, the same
approach can be used for the description of the 3-rd generation synchrotron
sources, however being mostly incoherent sources, especially in the
horizontal direction, they would require a large number of modes for a
sufficient description. This is illustrated in Fig. 9 where the ratio $\beta
_{j}/\beta _{0}$ of the eigenvalue $\beta _{j}$ to the lowest order
eigenvalue $\beta _{0}$ as a function of a mode number $j$ for the PETRA
III\ synchrotron source is presented. For comparison, results of the
calculations for the SASE1 undulator of the European XFEL are also shown in
the same figure.\ In the calculations we considered the same PETRA III
parameters as in the previous section (see Table \ref{table1}, high-$\beta $
operation of the PETRA III storage ring). Our results demonstrate that in
the vertical direction correlation functions can be properly described by
the contribution of eight modes (including the fundamental) and in the
horizontal direction a large number of modes (about 300) is necessary to
describe the coherence properties of the undulator source. This is in good
agreement with the behavior of the modes described by Eqs. (\ref{4.3a}, \ref%
{4.3b}) for a coherent and an incoherent source. Our results indicate that
in the vertical direction the undulator source is highly coherent, however
in the horizontal direction it behaves as an incoherent source.
\begin{figure}[tbp]
	\centering
	\includegraphics[width=0.6\textwidth]{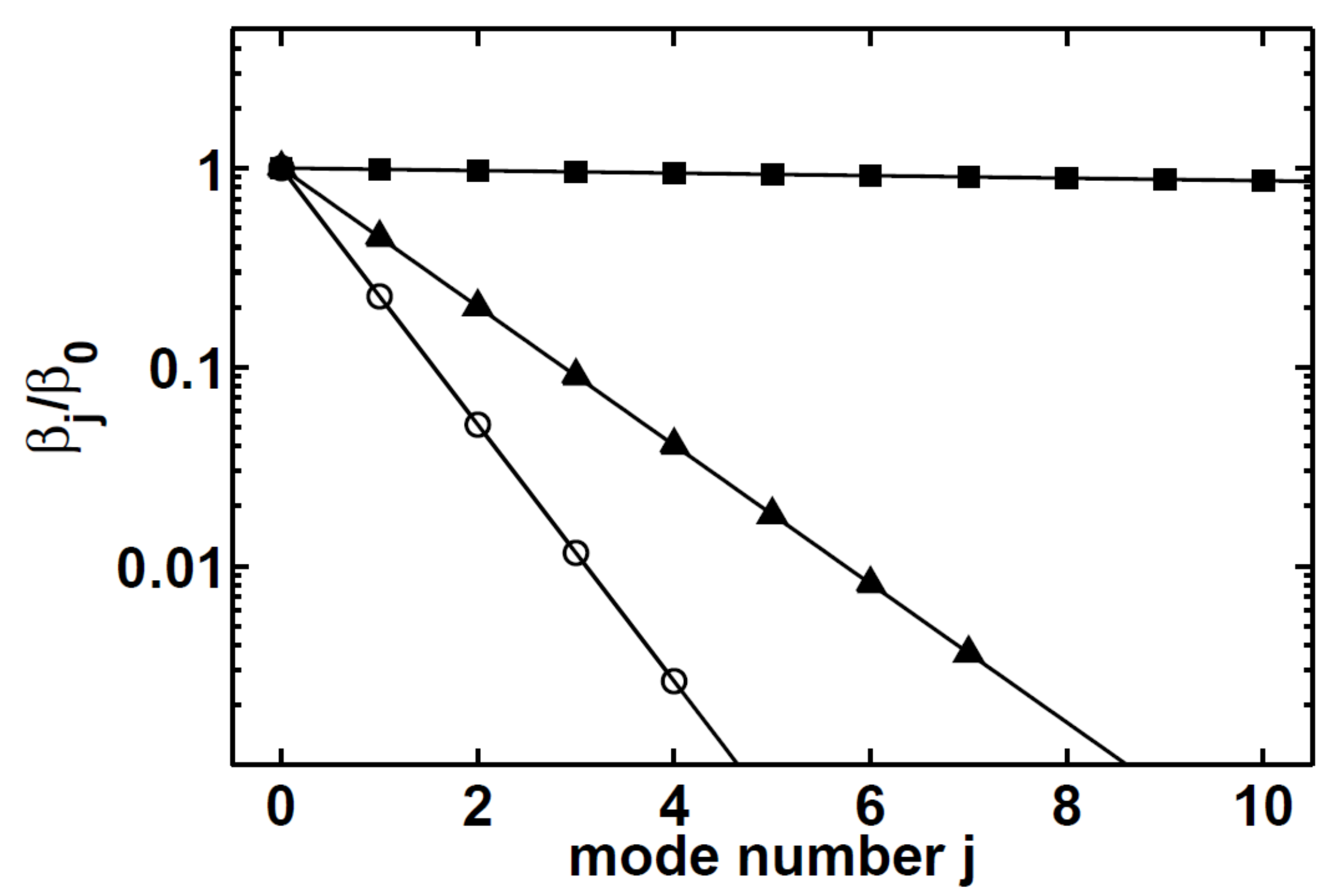} %
	\caption{The ratio $\beta _{j}/\beta _{0}$ of the eigenvalue $\beta
_{j}$ to the lowest order eigenvalue $\beta _{0}$ as a function of mode
number $j$. Results of the calculations for the parameters of the SASE1
undulator of the European XFEL (open circles), high-$\beta $ section of the
five meter undulator of the PETRA III source in the vertical direction
(triangles) and in the horizontal direction (squares). }
\end{figure}

\section{Summary}

In summary, we have demonstrated how a general theoretical approach based on
the results of statistical optics can be applied to give a sufficient
description of the correlation properties of the fields generated by
3-rd generation synchrotron sources and FELs. We have substituted a real source
by an equivalent planar GSM source with the same source size and divergence
as a real source. This phenomenological approach gives us the opportunity to
characterize this source with just two parameters, source size and
transverse coherence length. What is more important is that this approach
can be used as a tool to calculate correlation functions at different
distances from the source with simple analytic functions. In this way,
realistic estimates of the size of an x-ray beam and its coherence length
can be obtained at any distance from the source. This also gives a beam
profile and a complex degree of coherence in the transverse direction at any
distance from the source.

We applied this general approach to the concrete case of the PETRA III
source that is under construction. Our calculations have shown that the
coherence properties of this synchrotron source are quite different in the vertical and the
horizontal directions (this is typical of all 3-rd generation
synchrotron sources). In the vertical direction a beam produced with the
parameters of PETRA III is highly coherent with a degree of coherence
of about 40\%, however, in the horizontal direction it is a rather incoherent
source with a degree of coherence of about 1\%. Sixty meters downstream from
that source, where the first experimental hutches are planned, the transverse
coherence length in the vertical direction reaches a value of 190 $\mu $m and
in the horizontal direction can be about 25 $\mu $m. This has to be compared
with the size of the beam at the same distance with the FWHM values about
0.5 mm in the vertical and 1 mm in the horizontal directions. The values of
the transverse coherence length and the beam size scale linearly at these
large distances from the source and can be easily estimated at any other
distance from the source.

In the case of the XFELs we used a decomposition of the statistical fields
into a sum of independently propagating transverse modes for the analysis of
the coherence properties of these fields at different distances from the
source. Calculations were performed for the concrete case of the SASE1
undulator at the European XFEL, which is presently under construction. Our
analysis has shown that only a few transverse modes (five in the case of
European XFEL) contribute significantly to the total radiation field of the
XFEL. It was demonstrated that due to the contribution of a few transverse
modes, the SASE1 undulator source while being highly coherent (with the
degree of coherence about 60\%), can not be considered as fully coherent.
One essential difference between the radiation field from the XFEL compared
with that of a synchrotron source is that its coherence properties are
expected to be of the same order of magnitude in the vertical as well as in
the horizontal direction (compare with the results of the measurements
performed at FLASH \cite{SVK08}). The transverse coherence length 
500 m downstream from the source, where the first optical elements
will be located, is expected to be of the order of 350 $\mu $m compared to
the FWHM of the beam that is expected to be about 500 $\mu $m.

The approach used in this paper for the analysis of the transverse coherence
properties is quite general and can be applied as an effective and useful
tool for describing the coherence properties of undulator radiation at 3-rd
generation synchrotron sources and of SASE FELs. In our future work, we plan
to extend this approach to calculate the coherence properties of x-ray beams
passing through different optical elements.

  \renewcommand{\theequation}{A-\arabic{equation}}
  \setcounter{equation}{0}  
\renewcommand{\thefigure}{A\arabic{figure}}
  \setcounter{figure}{0}  

\renewcommand{\thetable}{A\arabic{table}}
  \setcounter{table}{0}  

 \section*{Appendix}  
We compared results obtained by the GSM with the results of Ref. \cite{GSS08}
for the PETRA III synchrotron source and different photon energies ranging
from 3 keV to 20 keV (see Figs. A1, A2). There are two critical
dimensionless parameters of the theory \cite{GSS08}%
\begin{equation}
D_{x,y}=k\sigma _{x,y}^{\prime 2}L_{u}\text{ and }N_{x,y}=\frac{k\sigma
_{x,y}^{2}}{L_{u}},  \label{A1}
\end{equation}%
where $\sigma _{x,y}$ and $\sigma _{x,y}^{\prime }$ are the electron bunch
sizes and divergences in the horizontal and in the vertical directions and $%
L_{u}$ is the undulator length. The electron beam sizes $\sigma _{x,y}$ and
divergences $\sigma _{x,y}^{\prime }$ can be calculated from the values of
the emittance $\varepsilon _{x,y}$ and known $\beta $-function of the
synchrotron source according to $\sigma _{x,y}=\sqrt{\varepsilon _{x,y}\beta
_{x,y}},\sigma _{x,y}^{\prime }=\sqrt{\varepsilon _{x,y}/\beta _{x,y}}$ (see
e.g. \cite{HK94}). As it was shown in Ref. \cite{GSS08}, for large
parameter values $D_{x,y}\gg 1$ and $N_{x,y}\gg 1$ the undulator source can
be described in the frame of the GSM (see Eqs. (\ref{3.1}, \ref{3.2})) with
the source size $\sigma _{Sx,y}=\sqrt{N_{x,y}L_{u}/k}$ and parameter $\delta
_{Sx,y}$ (\ref{3.3c}) $\delta _{Sx,y}=\sqrt{L_{u}/kD_{x,y}}$. Using the
values of $D_{x,y}$ and $N_{x,y}$ (\ref{A1}) and the definition of $\delta
_{Sx,y}$ (\ref{3.3c}) we obtain for the source size $\sigma _{Sx,y}=\sigma
_{x,y}$ and the coherence length $\xi _{Sx,y}$ at the source $\xi
_{Sx,y}=2\sigma _{x,y}/\sqrt{4k^{2}\varepsilon _{x,y}^{2}-1}$. These
expressions are exactly the same as discussed earlier in the paper (compare
for e.g. Eq. (\ref{3.11}) for the coherence length $\xi _{Sx,y}$) with the
only exception that here the electron beam parameters are used instead of
the photon beam parameters. For the PETRA III parameters at the photon
energy of $E=$12 keV and low-$\beta $ operation (see Table 1 and PETRA III
TDR \cite{PETRA}) we obtain $D_{x}=$234$,$ $D_{y}=$1.0 and $%
N_{x}=16, $ $N_{y}=$0.36. We can see from these estimates that in the
horizontal direction parameters $D_{x}\gg 1,N_{x}\gg 1$ that means that in
this direction GSM can be safely used. However, in the vertical direction at
the same energy $D_{y}\leq 1,$ $N_{y}\leq 1$, and a more careful analysis
has to be applied.

To compare predictions of the GSM theory with the theoretical results of
Ref. \cite{GSS08} in the vertical direction we used the far-field
expressions for the cross-spectral density function $W(\overline{y},\Delta
y) $ (Eq. (65) from Ref. \cite{GSS08}) and spectral density $S(y)$ (Eq. (71)
from Ref. \cite{GSS08}) obtained in the limit $D_{x}\gg 1$ and $N_{x}\gg 1$.
The SDC $\mu (\overline{y},\Delta y)$ was calculated according to Eq. (\ref%
{2.7}) 
\begin{equation}
\mu (\overline{y},\Delta y)=\frac{W(\overline{y},\Delta y)}{\sqrt{S(%
\overline{y}+\Delta y/2)}\sqrt{S(\overline{y}-\Delta y/2)}},  \label{A2}
\end{equation}%
where $\overline{y}=(y_{1}+y_{2})/2,\Delta y=y_{2}-y_{1}$ and $y_{1},$ $%
y_{2} $ are two positions in the vertical direction. 
Calculations were
performed for a five meter undulator of the PETRA III source, high-$\beta $
and\ low-$\beta $ operation, 60 m downstream from the source at the central
position of the beam ($\overline{y}=0)$. In Figs. A1 and A2 results of the
calculations of the SDC and spectral density for the energy range from 3 keV
up to 20 keV are presented (see for the parameters used in these simulations
Tables A1 and A2). These results are compared with the GSM \begin{figure}[!h]
	\centering
	\includegraphics[width=0.9\textwidth]{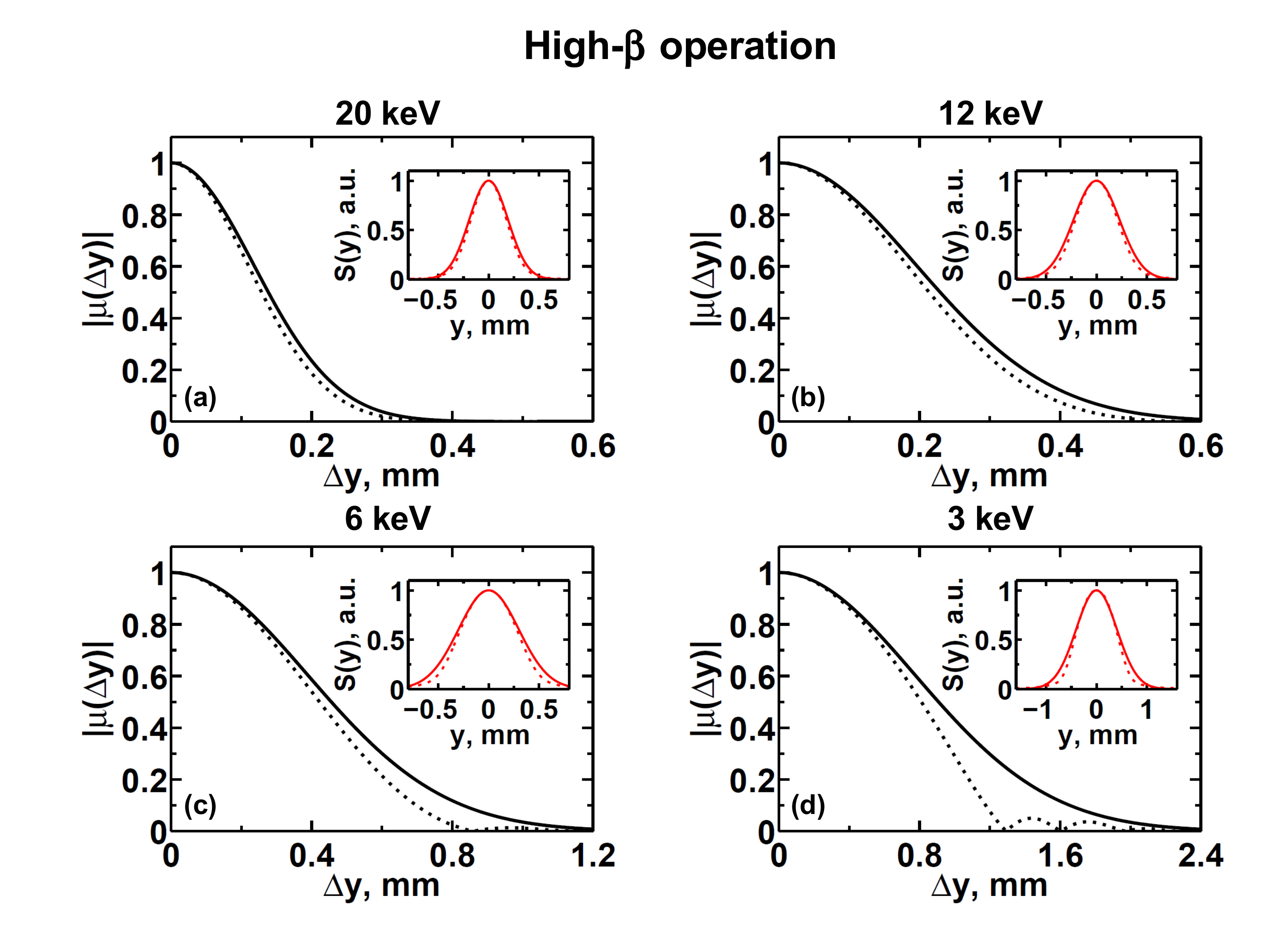} %
	\caption{ The absolute value of the spectral degree of coherence $%
\left\vert \mu (\Delta y)\right\vert $ in the vertical direction at the
distance 60 m downstream from the source for a high-$\beta $ operation
calculated for different photon energies ((a) 20 keV, (b) 12 keV, (c) 6 keV, (d) 3
keV) using results of Ref. \cite{GSS08} (dotted line). The spectral
density $S(y)$ (dotted line) calculated in the same conditions is shown in
the insets. For comparison, calculations performed in the frame of the GSM
are also shown in this figure (solid lines). }
\end{figure}
theory described
in this paper. For the GSM calculations at different photon energies the
total photon source size $\sigma _{Ty}$ and divergence $\sigma _{Ty}^{\prime
}$ were used. They are determined from a convolution of the sizes and
divergences of the electron beam ($\sigma _{y}$, $\sigma _{y}^{\prime }$)
with the intrinsic radiation characteristics of a single electron ($\sigma
_{r}$, $\sigma _{r}^{\prime }$). The latter are given by \cite{HK94} $\sigma
_{r}=\sqrt{2\lambda L_{u}}/4\pi ,\sigma _{r}^{\prime }=\sqrt{\lambda /2L_{u}}
$.
\begin{figure}[tbp]
	\centering
	\includegraphics[width=0.9\textwidth]{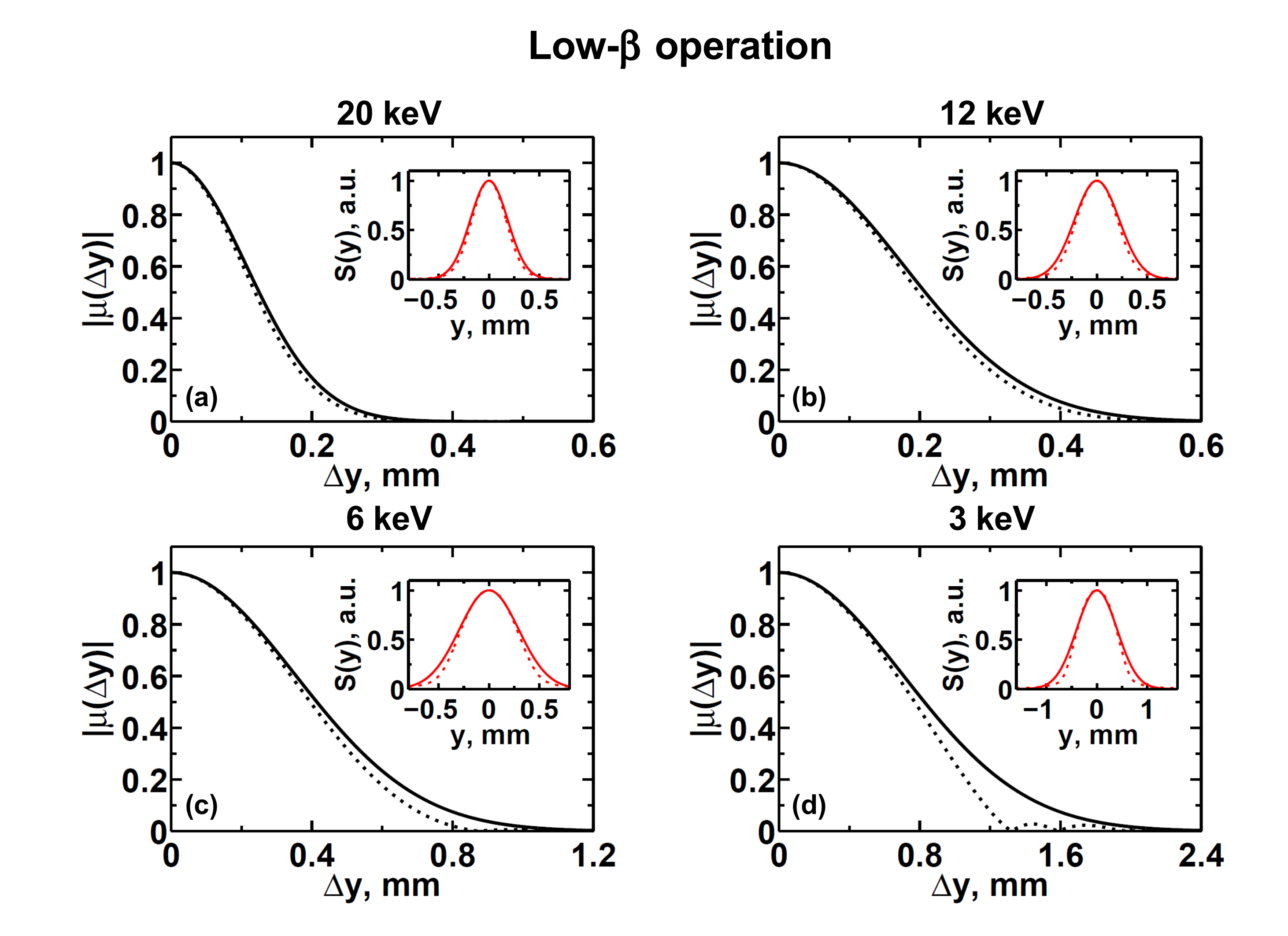} %
	\caption{The same as in Fig. A1 for a low-$\beta $ section of the
PETRA III storage ring. }
\end{figure}

As we can see from Figs. A1 and A2, in spite of the fact that parameters $%
D_{y}\leq 1,$ $N_{y}\leq 1$ (see Tables A1 and A2), the difference between
two approaches is negligible down to an energy of 6 keV. It becomes more
pronounced only at energies of about 3 keV for large separation distances $%
\Delta y$. It is also interesting to note that the coherence area, defined
as the area where the degree of coherence drops to 80\%, is the same in both
approaches down to a lowest energy of 3 keV. At the same time, for this very
low energy the effects of the single electron radiation (at the photon
energy $E=$3 keV $\sigma _{r}\geq \sigma _{y}$ and $\sigma _{r}^{\prime
}\geq \sigma _{y}^{\prime }$) are becoming more pronounced and reveal
themselves in the form of oscillations at large separations $\Delta y$. An
inspection of Figs. A1 and A2 shows that the GSM slightly overestimates the
values of the SDC compared to the results of Ref. \cite{GSS08}. We
relate this to the fact that at low energies, at the source position, the
intensity distribution obtained in the frame of the model \cite{GSS08}
contains long tails that effectively produce a larger source size in
comparison to a source size obtained by the GSM approach. We note
here as well that according to the Tables A1 and A2 the approximation $%
D_{x}\gg 1$ and $N_{x}\gg 1$ is no longer valid at very low energies
below 3 keV. Consequently, at these low energies Eqs. (65, 71) from Ref. 
\cite{GSS08} can not be applied for calculation of the coherence properties
of the five meter undulator source at PETRA III. A more careful treatment
using general expressions for the correlation functions should be used in
this case.

\begin{table}[!htb]
\centering          
\caption{Parameters of the synchrotron radiation source PETRA III
for a 5 m undulator, high-$\beta $ operation, and different photon
energies. Parameters $N_{x,y}$, $D_{x,y}$ are defined in Eq. (\ref{A1}), $
\sigma _{Tx,y}$ and $\sigma _{Tx,y}^{\prime }$ are the total photon source
sizes and divergences, $\sigma _{r}$ and $\sigma _{r}^{\prime }$ are the
intrinsic radiation characteristics of a single electron. The following
electron beam sizes $\sigma _{x}=141$ $\mu $m, $\sigma _{y}=4.9$ $\mu $m and
divergences $\sigma _{x}^{\prime }=7.1$ $\mu $rad, $\sigma _{y}^{\prime }=2.0
$ $\mu $rad were used in these calculations.}
\begin{tabular}{|c|c|c|c|c|}
	\multicolumn{2}{c}{ }\\
\hline
& 20 keV & 12 keV & 6 keV & 3 keV \\ \hline
$N_x$ & 405 & 243 & 122 & 61 \\ \hline
$D_x$ & 25 & 15 & 7.6 & 3.8 \\ \hline
$N_y$ & 0.5 & 0.30 & 0.15 & 0.07 \\ \hline
$D_y$ & 2.0 & 1.3 & 0.63 & 0.32 \\ \hline
$\sigma_{Tx},~\mu$m & 141 & 141 & 141 & 142 \\ \hline
$\sigma_{Tx}^{\prime},~\mu$rad & 7.5 & 7.7 & 8.4 & 9.6 \\ \hline
$\sigma_{Ty},~\mu$m & 5.3 & 5.5 & 6.1 & 7.1 \\ \hline
$\sigma_{Ty}^{\prime},~\mu$rad & 3.2 & 3.8 & 5.0 & 6.7 \\ \hline
$\sigma_r,~\mu$m & 2.0 & 2.6 & 3.6 & 5.1 \\ \hline
$\sigma_r^{\prime},~\mu$rad & 2.5 & 3.2 & 4.5 & 6.4 \\ \hline
\end{tabular}%
\label{table3}
\end{table}

\begin{table}[!htbp]
\centering   
\caption{The same as in Table A1 for the low-$\beta $ operation of
the synchrotron source. The following electron beam sizes $\sigma _{x}=36$ $%
\mu $m, $\sigma _{y}=5.5$ $\mu $m and divergences $\sigma _{x}^{\prime }=28$ 
$\mu $rad, $\sigma _{y}^{\prime }=1.8$ $\mu $rad were used here, $\sigma
_{r} $ and $\sigma _{r}^{\prime }$ are the same as in Table A1.
}
\begin{tabular}{|c|c|c|c|c|}
	\multicolumn{2}{c}{ }\\
\hline
& 20 keV & 12 keV & 6 keV & 3 keV \\ \hline
$N_x$ & 26 & 16 & 7.9 & 4.0 \\ \hline
$D_x$ & 390 & 234 & 117 & 58 \\ \hline
$N_y$ & 0.61 & 0.36 & 0.18 & 0.09 \\ \hline
$D_y$ & 1.7 & 1.0 & 0.51 & 0.25 \\ \hline
$\sigma_{T,x},~\mu$m & 36 & 36 & 36 & 36 \\ \hline
$\sigma_{T,x}^{\prime},~\mu$rad & 28 & 28 & 28 & 28 \\ \hline
$\sigma_{T,y},~\mu$m & 5.8 & 6.0 & 6.6 & 7.5 \\ \hline
$\sigma_{T,y}^{\prime},~\mu$rad & 3.1 & 3.7 & 4.9 & 6.7 \\ \hline
\end{tabular}%
\label{table4}
\end{table}

In conclusion, our analysis shows that for the high brilliance source PETRA
III the GSM can be safely used for the five meter undulator at x-ray
energies higher than 6 keV. It will also give a reasonable upper limit
estimate of the coherence length for the energies as low as 3 keV. Our
analysis has also shown that for a shorter undulator of 2 m length, which is typical for the
PETRA III source, both approaches give similar results even for lower
energies.\\


\textbf{Acknowledgments}

We acknowledge the help of M. Tischer in calculation of the radiation
properties of the 5 m PETRA III undulator using SRW code and the interest
and support of E. Weckert during the work on this project. We also
acknowledge a fruitful discussion with E. Saldin and M. Yurkov and careful
reading of the manuscript by A. Mancuso.


\begin{thebibliography}{99}
\bibitem{Footnote2} The applicability of the Gaussian Schell-model for the
description of the synchrotron radiation is discussed in details in Refs. 
\cite{THM98,GSS08,HK94,C95}.
\end{thebibliography}
\end{document}